\documentclass[11pt]{article}
\usepackage[utf8]{inputenc}
\usepackage{longtable}
\usepackage{subcaption}
\usepackage{graphicx}
\usepackage{placeins}
\usepackage{float}
\usepackage{amsmath, bbm, amssymb}
\usepackage[dvipsnames,table,xcdraw,svgnames]{xcolor}
\usepackage[hmargin=0.8in,vmargin={1.1in,1.1in}]{geometry}
\usepackage{hyperref}
\hypersetup{
   colorlinks=true,
    linkcolor=blue,
   citecolor=blue,      
   urlcolor=blue,
}
\usepackage{natbib}
\usepackage[font={footnotesize}]{caption}
\usepackage{url}

\usepackage{parskip}
\usepackage{array}
\newcolumntype{P}[1]{>{\centering\arraybackslash}p{#1}}

\title{Supply and demand shocks in the COVID-19 pandemic: \\
An industry and occupation perspective
\footnote{We would like to thank Eric Beinhocker, Stefania Innocenti, John Muellbauer, Marco Pangallo and David Vines for many comments and discussions. We are also grateful to Andrea Bacilieri and Luca Mungo for their help with the list of essential industries. We thank Baillie Gifford, IARPA, and the Oxford Martin School for the funding that made this possible.  \emph{Contacts:} rita.delriochanona@maths.ox.ac.uk, penny.mealy@inet.ox.ac.uk, anton.pichler@maths.ox.ac.uk, francois.lafond@inet.ox.ac.uk, doyne.farmer@inet.ox.ac.uk.} \footnote{$^\ddagger$These authors contributed equally.}
}

\author{R. Maria del Rio-Chanona ,$^{1, 2^{\ddagger }}$ Penny Mealy,$^{1,3,4^{\ddagger}}$  Anton Pichler,$^{1,2,6}$\\  Fran\c{c}ois Lafond,$^{1, 2}$ and 
 J. Doyne Farmer$^{1,2,5,6\dagger}$\\
\\
\footnotesize{$^{1}$ Institute for New Economic Thinking at the Oxford Martin School, University of Oxford}\\
\footnotesize{$^{2}$ Mathematical Institute, University of Oxford}\\
\footnotesize{$^{3}$ School of Geography and Environment, University of Oxford}\\
\footnotesize{$^4$ Bennett Institute for Public Policy, University of Cambridge}\\
\footnotesize{$^{5}$ Santa Fe Institute}\\
\footnotesize{$^6$ Complexity Science Hub Vienna}\\ 
\vspace{4mm}
\\
}
\date{\today}

\begin{document}

\maketitle

\begin{abstract}
We provide quantitative predictions of first order supply and demand shocks for the U.S. economy associated with the COVID-19 pandemic at the level of individual occupations and industries. To analyze the supply shock, we classify industries as essential or non-essential and construct a Remote Labor Index, which measures the ability of different occupations to work from home. Demand shocks are based on a study of the likely effect of a severe influenza epidemic developed by the US Congressional Budget Office. Compared to the pre-COVID period, these shocks would threaten around 22\% of the US economy’s GDP, jeopardise 24\% of jobs and reduce total wage income by 17\%. At the industry level, sectors such as transport are likely to have output constrained by demand shocks, while sectors relating to manufacturing, mining and services are more likely to be constrained by supply shocks. Entertainment, restaurants and tourism face large supply and demand shocks. At the occupation level, we show that high-wage occupations are relatively immune from adverse supply and demand-side shocks, while low-wage occupations are much more vulnerable. We should emphasize that our results are only first-order shocks -- we expect them to be substantially amplified by feedback effects in the production network. 


\end{abstract}
\newpage 

\newpage 
\section{Introduction}
The COVID-19 pandemic is having an unprecedented impact on societies around the world. As governments mandate social distancing practices and instruct non-essential businesses to close to slow the spread of the outbreak, there is significant uncertainty about the effect such measures will have on lives and livelihoods. While demand for specific sectors such as healthcare has skyrocketed in recent weeks, other sectors such as air transportation and tourism have seen demand for their services evaporate. At the same time, many sectors are experiencing issues on the supply-side, as governments curtail the activities of non-essential industries and workers are confined to their homes.  

Many economists and commentators believe that the economic impact could be dramatic \citep{baldwin2020intro}.  To give an example based on survey data in an economy under lockdown, the French statistical office estimated on March 26 that the economy is currently at around 65\% of its normal level\footnote{\url{https://www.insee.fr/en/statistiques/4473305?sommaire=4473307}}. \citet{bullard2020} provides an undocumented estimate that around a half of the US economy would be considered either essential, or able to operate without creating risks of diffusing the virus. \citet{inoue2020propagation} modeled how shutting down firms in Tokyo would cause a loss of output in other parts of the economy through supply chain linkages, and estimate that after a month, daily output would be 86\% lower than pre-shock (i.e. the economy would be operating at only $14\%$ of its capacity!). Using a calibrated extended consumption function, and assuming a labor income shock of 16\% and various consumption shocks by expenditure categories, \citet{muellbauer2020} estimates a fall of quarterly consumption of 20\%. Roughly speaking, most of these estimates, like ours, are estimates of instantaneous declines, and would translate to losses of annual GDP if the lockdown lasted for a year.

In a rare study based on aggregating industry-level shocks, the \citet{OECD2020evaluating} estimates a drop in immediate GDP of around 25\%, which is in line with our results. Another study by \citet{barrot2020} estimates industry level shocks by considering the list of essential industries, the closure of schools, and an estimate of the ability to work from home (based on ICT use surveys); Using these shocks in a multisector input-output model, they find that six weeks of social distancing would bring GDP down by 5.6\%.

In this paper, we aim to provide analytical clarity about the supply and demand shocks caused by public health measures and changes in preferences caused by avoidance of infection. We estimate (i) supply-side reductions due to the closure of non-essential industries and workers not being able to perform their activities at home and (ii) demand-side changes due to peoples' immediate response to the pandemic, such as increased demand for healthcare and reduced demand for goods or services that are likely to place people at risk of infection (e.g. tourism).

It is important to stress that the shocks that we predict here should not be interpreted as the \emph{overall impact} of the COVID-19 pandemic on the economy. Deriving overall impact estimates involves modeling second-order effects, such as the additional reductions in demand as workers who are stood down or laid off experience a reduction in income and additional reductions in supply as potential shortages propagate through supply chains. Further effects, such as cascading firm defaults, which can trigger bank failures and systemic risk in the financial system, could also arise.  Understanding these impacts requires a model of the macro-economy and financial sector.  We intend to present results from such an economic model in the future, but in the meantime we want to make our estimates of first order impacts available for researchers or governments to build upon or use in their own models.

Several researchers have already provided estimates of the supply shock from labor supply \citep{dingel2020,hicks2020,koren2020business}. Here we improve on these efforts in three ways: (i) we propose a methodology for estimating how much work can be done from home based on work activities, (ii) we identify industries for which working from home is irrelevant because the industries are considered essential, and (iii), we compare our estimated supply shocks to estimates of the demand shock, which in many industries is the more relevant constraint on output. 

To see why it is important to compare supply and demand shocks, consider the following thought experiment: Following social distancing measures, suppose industry $i$ is capable of producing only 70\% of its pre-crisis output, e.g. because workers can produce only 70\% of the output while working from home. If consumers reduce their demand by 90\%, the industry will produce only what will be bought, that is, 10\%. If instead consumers reduce their demand by 20\%, the industry will not be able to satisfy demand but will produce everything it can, that is, 70\%. In other words, the experienced first order reduction in output from the immediate shock will be the greater of the supply shock or the demand shock. In other words, most of the first-order impact on the economy will be due to an inability of people to work rather than to consume. However, again, we expect that as wages from work drop, there will be potentially larger second-order negative impacts on demand, and the potential for a self-reinforcing  downward spiral in output, employment, income, and demand.

Overall, we find that the supply and demand shocks considered in this paper represent a reduction of around one quarter of the US economy's value added, one fifth of current employment and about 17\% of the US total wage income. Supply shocks account for the majority of this reduction. These effects vary substantially across different industries. While we find no negative effects on value added for industries like Legal services, Power generation and distribution or Scientific research, the expected loss of value added reaches up to 80\% for Accommodation, Food services and Independent artists.

We show that sectors such as Transports are likely to experience immediate demand-side reductions that are larger than their corresponding supply-side shocks. Other industries such as manufacturing, mining and certain service sectors are likely to experience larger immediate supply-side shocks relative to demand-side shocks. Health unsurprisingly experiences an overall increase in demand for its output. Entertainment, restaurants and hotels experience very large supply and demand shocks, with the demand shock dominating. These results are important because supply and demand shocks might have different degrees of persistence, and industries will react differently to policies depending on the constraints that they face. Overall, however, we find that aggregate effects are dominated by supply shocks, with a large part of manufacturing and services being classified as non-essential while its labor force is unable to work from home.

We also break down our results by occupation and show that there is a strong negative relationship between the overall immediate shock experienced by an occupation and its wage. Relative to the pre-COVID period, 38\% of the jobs for workers in the bottom quartile of the wage distribution are predicted to be vulnerable. (And bear in mind that this is only a first-order shock -- second order shocks may significantly increase this). In contrast, most high-wage occupations are relatively immune from adverse shocks, with only 6\% of the jobs at risk for the 25\% of workers working in the highest pay occupations. Absent strong support from governments, most of the economic burden of the pandemic will fall on lower wage workers.

We neglect several effects that, while important, are small compared to those we consider here.  First, we have not sought to quantify the reduction in labor supply due to workers contracting COVID-19. A rough estimate suggests that this effect is relatively minor in comparison to the shocks associated with social distancing measures that are being taken in most developed countries.\footnote{
See Appendix \ref{appendix:epidemiology} for rough quantitative estimates in support of this argument.
}
We have also not explicitly included the effect of school closures.  However, in Appendix \ref{appendix:school} we argue that this is not the largest effect and is already partially included in our estimates through indirect channels.   

A more serious problem is caused by the need to assume that that within a given occupation, being unable to perform some work activities does not harm the performance of other work activities. Within an industry, we also assume that if workers in a given occupation cannot work, they do not produce output, but this does not prevent other workers in different occupations from producing.  In both cases we assume that the effects of labor on production are linear, i.e. that production is proportional to the fraction of workers who can work. In reality however, it is clear that there are important complementarities leading to nonlinear effects.  There are many situations where production requires a combination of different occupations, such that if workers in key occupations cannot work at home, production is not possible. For example, while the accountants in a steel plant might be able to work from home, if the steelworkers needed to run the plant cannot come to work, no steel is made.  We cannot avoid making linear assumptions because as far as we know there is no detailed understanding of the labor production function and these interdependencies at an industry level.  By neglecting nonlinear effects, our work here should consequently be regarded as an approximate lower bound on the size of the first order shocks.

This paper focuses on the United States. We have chosen it as our initial test case because input-output tables are more disaggregated than those of most other countries, and because the O*NET database, which we rely on for information about occupations was developed based on US data. With some additional assumptions it is possible to apply the analysis we perform here to other developed countries.

This paper is structured as follows. In Section \ref{sec:supply_shock} we describe our methodology for estimating supply shocks, which involves developing a new Remote-Labor Index (RLI) for occupations and combining it with a list of essential industries. Section \ref{sec:demand_shock} discusses likely demand shocks based on estimates developed by the US \citet{CBO2006} to predict the potential economic effects of an influenza pandemic. In Section \ref{sec:combining_supply_demand}, we show a comparison of the supply and demand shocks across different industries and occupations and identify the extent to which different activities are likely to be constrained by supply or demand. In this section, we also explore which occupations are more exposed to infection and make comparisons to wage and occupation-specific shocks. Finally, in Section \ref{sec:discussion} we discuss our findings in light of existing research and outline avenues for future work. We also make all of our data available in a continuously updated online repository.

\section{Supply shocks}
\label{sec:supply_shock}

Supply shocks from pandemics are mostly thought of as labor supply shocks. 
Several pre-COVID-19 studies focused on the direct loss of labor from death and sickness (e.g. \citet{mckibbin2006global, santos2013risk}), although some have also noted the potentially large impact of school closure \citep{keogh2010possible}. \citet{mckibbin2020global} consider (among other shocks) reduced labor supply due to mortality, morbidity due to infection, and morbidity due to the need to care for affected family members. In countries where social distancing measures are in place, social distancing measures will have a much larger economic effect than the direct effects from mortality and morbidity.  This is in part because if social distancing measures work, only a small share of population will be infected and die eventually. Appendix \ref{appendix:epidemiology} provides more quantitative estimates of the direct mortality and morbidity effects and argues that the they are likely to be at least an order of magnitude smaller than those due to social distancing measures, especially if the pandemic is contained.

\begin{figure}[!ht]
    \centering
\includegraphics[page = 2, width = \textwidth, trim={0cm 2.5cm 0 2.5cm},clip]{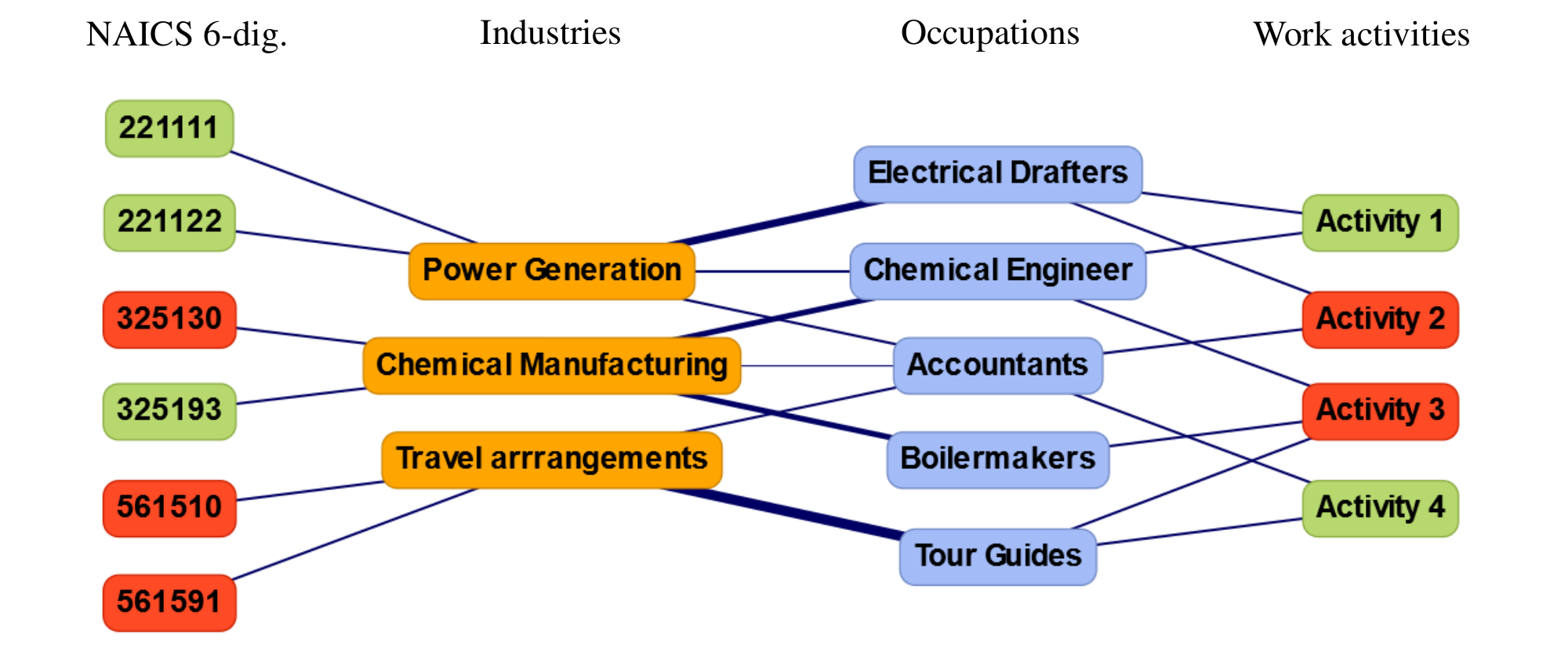}
    \caption{A schematic network representation of supply side shocks. The nodes to the left represent the list of essential industries at the NAICS 6-digit level. A green node indicates essential, a red node non-essential.
    The orange nodes (center-left) are more aggregate industry categories (e.g. 4-dig. NAICS or the BLS industry categories) for which further economic data is available.
    These two sets of nodes are connected through industry concordance tables.
    The blue nodes (center-right) are different occupations. A weighted link connecting an industry category with an occupation represents the number of people of a given occupation employed in each industry.
    Nodes on the very right are O*NET work activities. Green work activities mean that they can be performed from home, while red means that they cannot. O*NET provides a mapping of work activities to occupations.
    }
        \label{fig:supply_shocks_main}
\end{figure}

For convenience we neglect mortality and morbidity and assume that the supply shocks are determined only by the amount of labor that is withdrawn due to social distancing. We consider two key factors: (i) the extent to which workers in given occupations can perform their requisite activities at home and (ii) the extent to which workers are likely to be unable to come to work due to being in non-essential industries. We quantify these effects on both industries and occupations.  Figure \ref{fig:supply_shocks_main} gives a schematic overview of how we predict industry and occupation specific supply shocks. We explain this in qualitative terms in the next few pages; for a formal mathematical description see Appendix \ref{apx:supply_shocks}.

\subsection{How much work can be performed from home?}
One way to assess the degree to which workers are able to work from home during the COVID-19 pandemic is by direct survey. For example, \citet{zhang2020health} conducted a survey of  Chinese citizens in late February (one month into the coronavirus-induced lockdown in China) and found that 27\% of the labor force continued  working at the office, 38\% worked from home, and 25\% stopped working. \citet{adams2020} surveyed US and UK citizens in late March, and reported that the share of tasks that can be performed from home varies widely between occupations (from around 20 to 70\%), and that higher wage occupations tend to be more able to work from home. 

Other recent work has instead drawn on occupation-level data from the Occupational Information Network (O*NET) to assess `work context' characteristics to determine  labour shocks due to the COVID-19 pandemic. For example, \cite{hicks2020} considered the degree to which an occupation is required to `work with others' or involves `physical proximity to others' in order to assess which occupations are likely to be most impacted by social distancing. Similarly, to identify which occupations were likely to be able to work from home, \cite{dingel2020} used O*NET data on the extent to which occupations requires daily `work outdoors' or needed to engage with `operating vehicles, mechanized devices or equipment'. 

We go to a more granular level than `work context', and instead draw on O*NET's `intermediate work activity' data, which provides a list of the activities performed by each occupation based on a list of 332 possible work activities. For example, a Nurse undertakes activities such as “maintain health or medical records”, “develop patient or client care or treatment plans” and “operate medical equipment’, while a Computer Programmer performs activities such as “resolve computer programs”, “program computer systems or production equipment” and “document technical designs, producers or activities”\footnote{
In the future we intend to redo this using O*NET’s “detailed" work activity data, which involves over 2000 individual activities associated with different occupations.  We believe this would somewhat improve our analysis, but we also think that for our purposes here the intermediate activity list provides a good approximation. }.  
In Fig \ref{fig:supply_shocks_main} these work activities are illustrated by the rightmost set of nodes.

\paragraph{\textit{Which work activities can be performed from home?}}

Four of us independently assigned a subjective binary rating to each work activity as to whether it could successfully be performed at home. The individual results were in broad agreement. Based on the responses, we assigned an overall consensus rating to each work activity\footnote{
An activity was considered to be able to performed at home if three or more respondents rated this as true. We also undertook a robustness analysis where an activity was considered to able to be performed at home based on two or more true ratings. Results remained fairly similar. In post-survey discussion, we agreed that the most contentious point is that some work activities might be done from home or not, depending on the industry in which it is performed.}.
Ratings for each work activity are available in an online data repository\footnote{
\url{https://zenodo.org/record/3751068}}.
While O*NET maps each intermediate work activity to 6-digit O*NET occupation codes, employment information from the US Bureau of Labor Statistics (BLS) is available for the 4-digit 2010 Standard Occupation Scheme (SOC) codes, so we mapped O*NET and SOC codes using a crosswalk available from O*NET.\footnote{
Available at \url{https://www.onetcenter.org/crosswalks.html}.
} Our final sample contains 740 occupations.

\begin{table}[H]
\small
    \centering
    \begin{tabular}{|l|l|}
    \hline
        \textbf{Occupation} & \textbf{RLI} \\ \hline
    \hline
        Credit Analysts & 1.00 \\ \hline
        Insurance Underwriters & 1.00 \\ \hline
        Tax Preparers & 1.00 \\ \hline
        Mathematical Technicians & 1.00 \\ \hline
        Political Scientists & 1.00 \\ \hline
        Broadcast News Analysts & 1.00 \\ \hline
        Operations Research Analysts & 0.92 \\ \hline
        Eligibility Interviewers, Government Programs & 0.92 \\ \hline
        Social Scientists and Related Workers, All Other & 0.92 \\ \hline
        Technical Writers & 0.91 \\ \hline
        Market Research Analysts and Marketing Specialists & 0.90 \\ \hline
        Editors & 0.90 \\ \hline
        Business Teachers, Postsecondary & 0.89 \\ \hline
        Management Analysts & 0.89 \\ \hline
        Marketing Managers & 0.88 \\ \hline
        Mathematicians & 0.88 \\ \hline
        Astronomers & 0.88 \\ \hline
        Interpreters and Translators & 0.88 \\ \hline
        Mechanical Drafters & 0.86 \\ \hline
        Forestry and Conservation Science Teachers, Postsecondary & 0.86 \\ \hline

        \dots & \dots \\ \hline
         Bus and Truck Mechanics and Diesel Engine Specialists & 0.00 \\ \hline
        Rail Car Repairers & 0.00 \\ \hline
        Refractory Materials Repairers, Except Brickmasons & 0.00 \\ \hline
        Musical Instrument Repairers and Tuners & 0.00 \\ \hline
        Wind Turbine Service Technicians & 0.00 \\ \hline
        Locksmiths and Safe Repairers & 0.00 \\ \hline
        Signal and Track Switch Repairers & 0.00 \\ \hline
        Meat, Poultry, and Fish Cutters and Trimmers & 0.00 \\ \hline
        Pourers and Casters, Metal & 0.00 \\ \hline
        Foundry Mold and Coremakers & 0.00 \\ \hline
        Extruding and Forming Machine Setters, Operators, and Tenders, Synthetic and Glass Fibers & 0.00 \\ \hline
        Packaging and Filling Machine Operators and Tenders & 0.00 \\ \hline
        Cleaning, Washing, and Metal Pickling Equipment Operators and Tenders & 0.00 \\ \hline
        Cooling and Freezing Equipment Operators and Tenders & 0.00 \\ \hline
        Paper Goods Machine Setters, Operators, and Tenders & 0.00 \\ \hline
        Tire Builders & 0.00 \\ \hline
        Helpers--Production Workers & 0.00 \\ \hline
        Production Workers, All Other & 0.00 \\ \hline
        Machine Feeders and Offbearers & 0.00 \\ \hline
        Packers and Packagers, Hand & 0.00 \\ \hline
    \end{tabular}
        \caption{{\bf Top and bottom 20 occupations ranked by Remote Labor Index (RLI)}, based on proportion of work activities that can to be performed by home. There are 44 occupations with an RLI of zero; we show only a random sample.}
    \label{tab:rli}
\end{table}

\paragraph{\textit{From work activities to occupations.}}
We then created a Remote Labor Index (RLI) for each occupation by calculating the proportion of an occupation’s work activities that can be performed at home.
An RLI of 1 would indicate that all of the activities associated with an occupation could be undertaken at home, while an RLI of 0 would indicate that none of the occupation's activities could be performed at home.\footnote{
We omitted ten occupations that had less than five work activities associated with them. These occupations include Insurance Appraisers Auto Damage; Animal Scientists; Court Reporters; Title Examiners, Abstractors, and Searchers; Athletes and Sports Competitors; Shampooers; Models; Fabric Menders, Except Garment; Slaughterers and Meat Packers' and Dredge Operators.}
%
The resulting ranking of each of the 740 occupations can be found in the online repository (see footnote 5).  While the results are not perfect\footnote{
There are a few cases that we believe are misclassified.  For example, two occupations with a high RLI that we think cannot be performed remotely are real estate agents (RLI = 0.7) and retail salespersons (RLI = 0.63). 
However, these are exceptions -- in most cases the rankings make sense.  The full list can be examined on our online repository at \url{https://zenodo.org/record/3751068}.  We believe the problems will be fixed when we redo the analysis using fine grained work activities, and we doubt that our results will be qualitatively changed.},
most of the rankings make sense.  For example, in Table \ref{tab:rli}, we show the top 20 occupations having the highest RLI ranking. Some occupations such as credit analysts, tax preparers and mathematical technician occupations are estimated to be able to perform $100\%$ of their work activities from home. Table \ref{tab:rli} also shows a sample of the 43 occupations with an RLI ranking of zero, i.e. those for which there are no activities that are able to be performed at home.

\begin{figure}[H]
    \centering
\includegraphics[width = 0.8\textwidth]{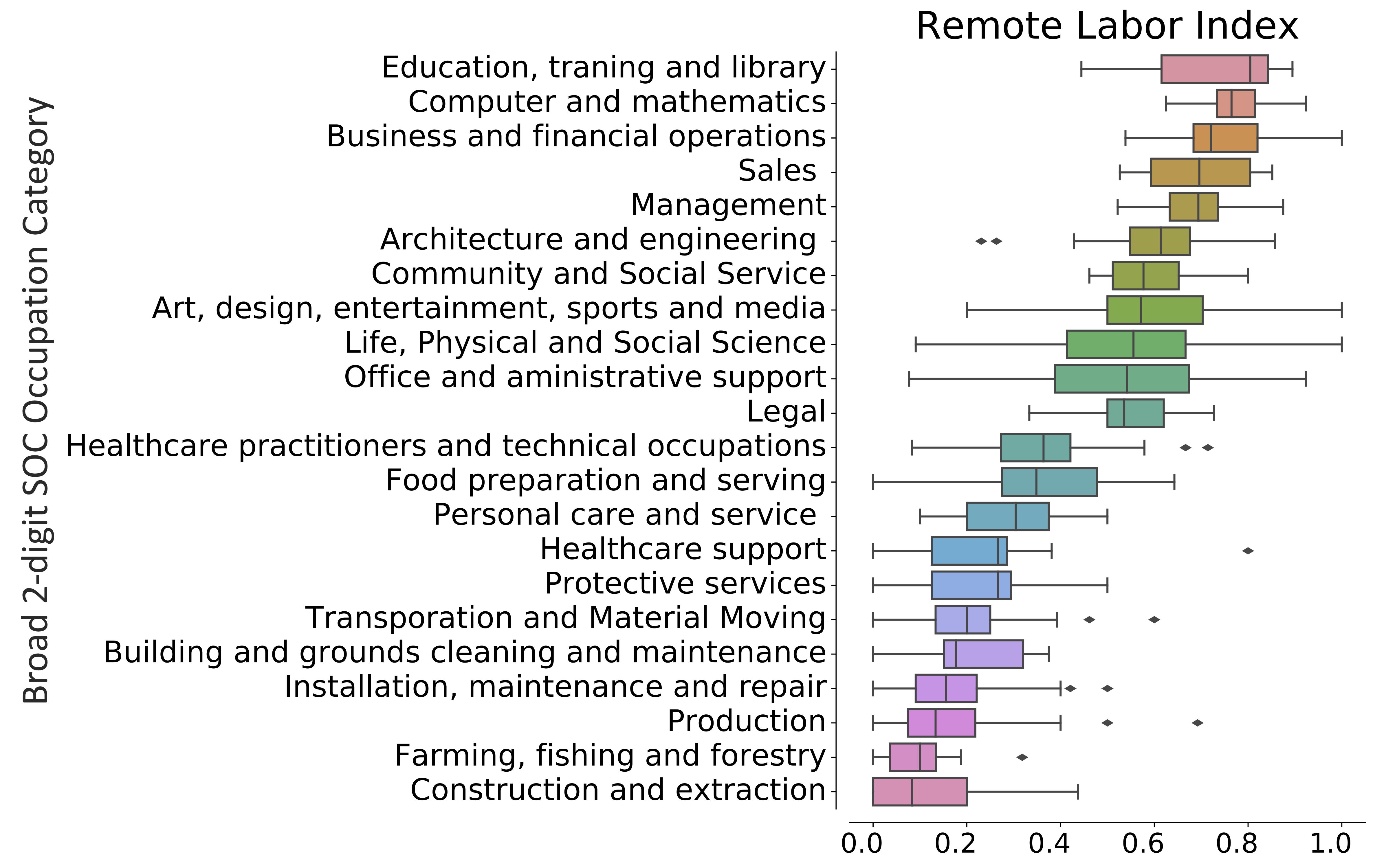}
    \caption{{\bf Distribution of Remote Labor Index across occupations}.  We provide boxplots showing distribution of RLI for each 4-digit occupation in each 2-digit SOC occupation category.}
        \label{fig:occ_broad_RLI_boxplot}
\end{figure}

To provide a broader perspective of how the RLI differs across occupation categories, Figure \ref{fig:occ_broad_RLI_boxplot} shows a series of box-plots indicating the distribution of RLI for each 4-digit occupation in each 2-digit SOC occupation category. We have ordered 2-digit SOC occupations in accordance with their median values. Occupations with the highest RLI relate to Education, training and library, Computer and Mathematical, and Business and Financial roles, while occupations relating to Production, Farming, Fishing and Forestry, and Construction and Extraction tend to have lower RLI.

\paragraph{\textit{From occupations to industries.}}
We next map the RLI to industry categories to quantify industry-specific supply shocks from social distancing measures. We obtain occupational compositions per industry from the BLS, which allows us to match 740 occupations to 277 industries\footnote{
We use the May 2018 Occupational Employment Statistics (OES) estimates on the level of 4-digit NAICS (North American Industry Classification System), file \textit{nat4d\_M2018\_dl}, which is available at \url{https://www.bls.gov/oes/tables.htm} under \textit{All Data}.
Overall, our merged dataset covers 136.8 out of 144 million employed people (95\%) initially reported in the OES.
}.

\begin{figure}[H]
    \centering
\includegraphics[width = 0.8\textwidth]{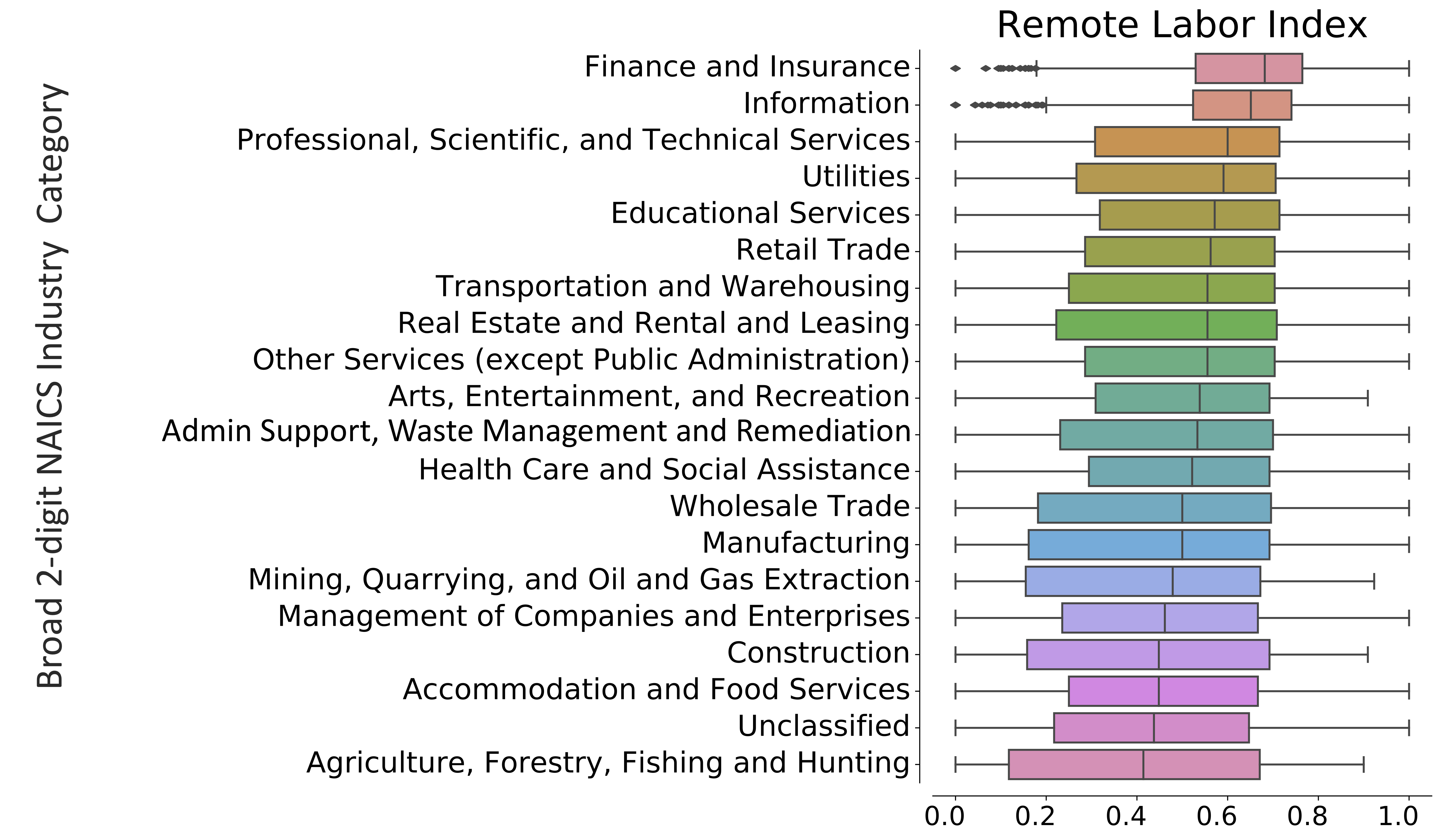}
    \caption{{\bf Distribution of Remote Labor Index across industries}.  We provide boxplots showing distribution of RLI for each 4-digit occupation in each 2-digit NAICS Industry category.}
        \label{fig:ind_broad_RLI_boxplot}
\end{figure}

In Figure \ref{fig:ind_broad_RLI_boxplot}, we show the RLI distribution for each 4-digit occupation category falling within each broad 2-digit NAICS category. Similar to Figure \ref{fig:occ_broad_RLI_boxplot}, we have ordered the 2-digit NAICS industry categories in accordance with the median values of each underpinning distribution. As there is a greater variety of different types of occupations within these broader industry categories, distributions tend to be much wider. Industries with the highest median RLI values relate to Information, Finance and Insurance, and Professional, Science and Technical Services, while industries with the lowest median RLI relate to Agriculture, Forestry, Fishing and Hunting and Accommodation and Food Services. 

In the Appendix, we show industry-specific RLI values for the more detailed 4-digit NAICS industries. To arrive at a single number for each 4-digit industry, we compute the employment-weighted average of occupation-specific RLIs.
The resulting industry-specific RLI can be interpreted as a rough estimate of the fraction of jobs which can be performed from home for each industry.

\subsection{Which industries are ``essential''?}

Across the world, many governments have mandated that certain industries deemed `essential’ should remain open over the COVID-19 crisis duration. What constitutes an ‘essential’ industry has been the subject of significant debate, and it is likely that the endorsed set of essential industries will vary across countries. As the U.S. government has not  produced a definitive list, here we draw on the list of essential industries developed by Italy and assume it can be applied, at least as an approximation, to other countries such as the U.S. as well.  This list has two key advantages. First, as Italy was one of the countries affected earliest and most severely, it was one of the first countries to invest significant effort considering which industries should be deemed essential. Second, Italy’s list of essential industries includes NACE industrial classification codes, which can be mapped to the NAICS industry classification we use to classify industrial employment in this paper.\footnote{Mapping NACE industries to NAICS industries is not straightforward. NACE industry codes at the 4-digit level are internationally defined. However, 6-digit level NACE codes are country specific.  Moreover, the list of essential industries developed by Italy involves industries defined by varying levels of aggregation. Most essential industries are defined at the NACE 2-digit and 4-digit level, with a few 6-digit categories thrown in for good measure. As such, much of our industrial mapping methodology involved mapping from one classification to the other by hand. We provide a detailed description of this process in Appendix \ref{apx:nace-naicsmap}.} 

Table \ref{tab:naics_essential} shows the total numbers of NAICS essential industries at the 6-digit and 4-digit level. More than 50\% of 6-digit NAICS industries are considered essential. At the 6-digit level the industries are either classified as essential, and assigned essential score  $1$, or non-essential and assigned essential score $0$. Unfortunately, it is not possible to translate this directly into a labour force proportion as BLS employment data at detailed occupation and industry levels are only available at the NAICS 4-digit level. To derive an estimate at the 4-digit level, we assume that labor in a NAICS 4-digit code is uniformly distributed over its associated 6-digit codes. We then assign an essential `share' to each 4-digit NAICS industry based on the proportion of its 6-digit NAICS industries that are considered essential. (The distribution of the essential share over 4-digit NAICS industries is shown in the Appendix). Based on this analysis, we estimate that about 89 million (or 64\%) of US workers are currently employed in essential industries. 

\begin{table}[!ht]
    \centering
    \begin{tabular}{|l|l|}
    \hline
        Total 6-digit NAICS industries & 1057 \\ 
    Number of essential 6-digit NAICS industries & 612 \\ 
        Fraction of essential industries at 6-digit NAICS & 0.58 \\
        \hline
        Total 4-digit NAICS industries in our sample & 277 \\ 
    Average rating of essential industries at 4-digit NAICS & 0.56 \\ 
        Fraction of labor force in essential industries & 0.68 \\ 
        \hline
    \end{tabular}
    \caption{\textbf{Essential industries.} Essential industries at the 6-digit level and essential `share' at the 4-digit level.  Note that 6-digit NAICS industry classifications are binary (0 or 1) whereas 4-digit NAICS industry classifications can take on any value between 0 and 1. }
    \label{tab:naics_essential}
\end{table}

\subsection{Supply shock: non-essential industries unable to work from home}

\begin{figure}[H]
    \centering
\includegraphics[width = 1.0\textwidth]{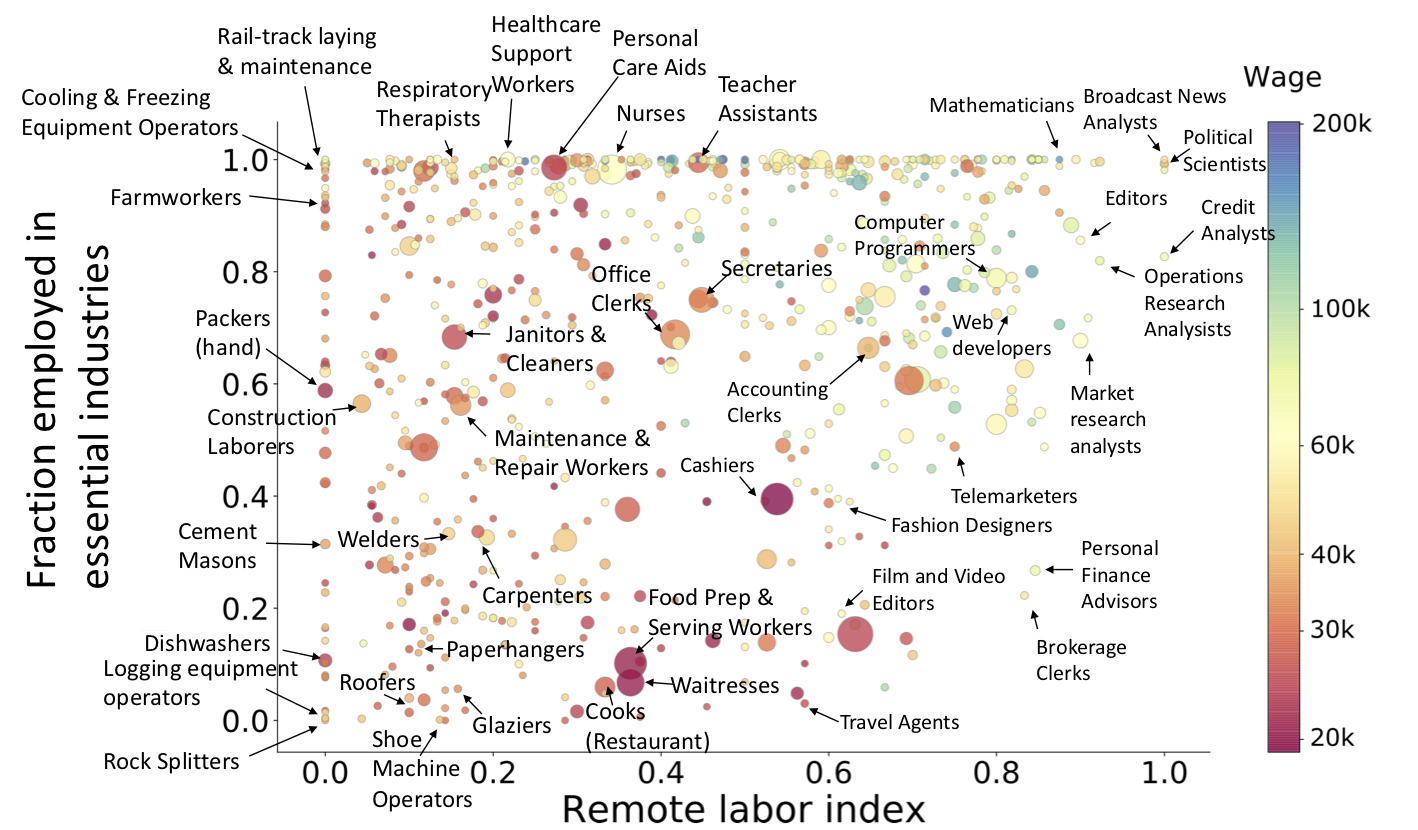}
    \caption{{\bf Fraction employed in an essential industry vs Remote Labor Index for each occupation}. Omitting the effect of demand reduction, the occupations in the lower left corner, with a small proportion of workers in essential industries and a low Remote Labor Index, are the most vulnerable to loss of employment due to social distancing.}
    \label{fig:occ_ind_broad_panel}
\end{figure}

Having analyzed both the extent to which jobs in each industry are essential and the likelihood that workers in a given occupation can perform their requisite activities at home, we now combine these to consider the overall first-order effect on labor supply in the US. In Figure  \ref{fig:occ_ind_broad_panel}, we plot the Remote Labor Index of each occupation against the fraction of that occupation employed in an essential industry. Each circle in the scatter plot represents an occupation; the circles are sized proportional to current employment and color coded according to the median wage in each occupation. 

Figure \ref{fig:occ_ind_broad_panel} indicates the vulnerability of occupations due to supply-side shocks. Occupations in the lower left-hand side of the plot (such as Diswashers, Rock Splitters and Logging Equipment Operators) have lower RLI scores (indicating they are less able to work from home) and are less likely to be employed in an essential industry. If we consider only the immediate supply-side effects of social distancing, workers in these occupations are more likely to face reduced work hours or be at risk of losing their job altogether. In contrast, occupations on the upper right-hand side of the plot, such as Credit Analysis, Political Scientists and Operations Research Analysts) have higher RLI scores and are more likely to employed in an essential industry. These occupations are less economically vulnerable to the supply-side shocks (though, as we discuss in the next section, they could still face employment risks due to first-order demand-side effects). Occupations in the upper-left hand side of the plot (such as Farmworkers, Healthcare Support Workers and Respiratory Therapists) are less likely to be able to perform their job at home, but since they are more likely to be employed in an essential industry their economic vulnerability from supply-side shocks is lower. Interestingly, there are relatively few occupations on the lower-right hand side of the plot. This indicates that occupations that are predominantly employed in non-essential industries tend to be less able to perform their activities at home.

\begin{figure}[H]
    \centering
\includegraphics[width = 0.5\textwidth]{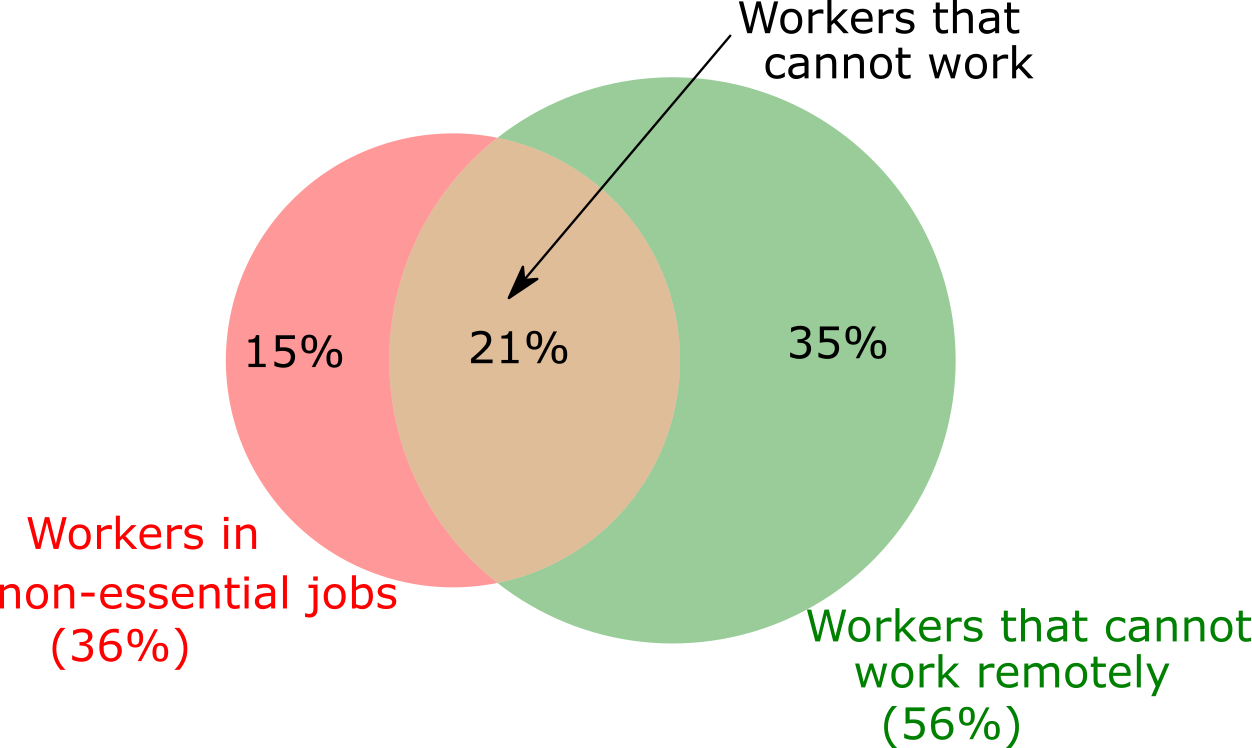}
    \caption{{\bf Workers that cannot work}. On the left is the percentage of workers in a non-essential job ($36\%$ in total). On the right is the percentage of workers that cannot work remotely ($56\%$ in total). The intersection is the set of workers that cannot work, which is $19\%$ of all workers. A remaining 29\% of workers are in essential jobs where they can work remotely.}
        \label{fig:venn_ess_remote}
\end{figure}

To help visualize the problem we provide a summary in the form of a Venn diagram in Figure~\ref{fig:venn_ess_remote}. Before the pandemic, 36\% of workers were employed in non-essential jobs. 56\% of workers cannot do their job remotely.  21\% of workers are in the intersection corresponding to non-essential jobs that cannot be performed remotely.  In addition, there are 29\% of workers in essential industries that can also work from home.\footnote{
In fact we allow for a continuum between the ability to work from home, and an industry can be partially essential.}
\section{Demand shock}\label{sec:demand_shock}

The pre-COVID-19 literature on epidemics and the discussions of the current crisis make it clear that epidemics strongly influence patterns of consumer spending. In addition to increasing their demand for health services, consumers are also likely to seek to reduce their risk of exposure to the virus and decrease demand for products and services that involve close contact with others. In the early days of the outbreak, stockpiling behaviour also drives a direct demand increase in the retail sector \citep{baker2020how}.

\paragraph{Estimates from the CBO.} Our estimates of the demand shock are based on expert estimates developed by the US \citet{CBO2006} that attempted to predict the potential impact of an influenza pandemic. Similar to the current COVID-19 pandemic, this analysis assumes that demand is reduced due to the desire to avoid infection. While the analysis is highly relevant to the present COVID-19 situation, 
it is important to note that the estimates are ``extremely rough'' and ``based loosely on Hong Kong’s experience with SARS''. The CBO provides two scenarios. We take the severe scenario, which ``describes a pandemic that is similar to the 1918-1919 Spanish flu outbreak. It incorporates the assumption that a particularly virulent strain of influenza infects roughly 90 million people in the United States and kills more than 2 million of them''.

In this paper, we simply take the CBO estimates as immediate (first-order) demand-side shocks. The CBO lists demand side estimates for broad industry categories, which we mapped to the 2-digit NAICS codes by hand. Table \ref{table:cbo} shows the CBO's estimates of the percent decrease in demand by industry, and Table \ref{table:cbonaics} in Appendix \ref{appendix:demand} shows the full mapping to 2-digit NAICS. 

\begin{table}[ht]
\centering
\begin{tabular}{|l|l|}
  \hline
Broad industry name & Severe Scenario Shock \\ 
  \hline
Agriculture & -10 \\ 
  Mining & -10 \\ 
  Utilities & 0 \\ 
  Construction & -10 \\ 
  Manufacturing & -10 \\ 
  Wholesale trade & -10 \\ 
  Retail trade & -10 \\ 
  Transportation and warehousing (including air, rail and transit) & -67 \\ 
  Information (Published, broadcast) & 0 \\ 
  Finance & 0 \\ 
  Professional and business services & 0 \\ 
  Education & 0 \\ 
  Healthcare & 15 \\ 
  Arts and recreation & -80 \\ 
  Accommodation/food service & -80 \\ 
  Other services except government & -5 \\ 
  Government & 0 \\ 
   \hline
\end{tabular}
\caption{Demand shock by sector according to the Congressional Budget Office (2006)'s severe scenario.} 
\label{table:cbo}
\end{table}

These estimates, of course, are far from perfect. They are based on expert estimates made more than ten years ago for an hypothetical pandemic scenario. It is not entirely clear if they are for gross output or for final (consumer) demand. However, in Appendix \ref{appendix:demand}, we describe three other sources of consumption shocks \citep{keogh2010possible,muellbauer2020,OECD2020evaluating} that provide broadly similar estimates by industry or spending category. 

\paragraph{Transitory and permanent shocks.} An important question is whether demand reductions are just postponed expenses, and if they are permanent \citep{mann2020,keogh2010possible}. \cite{baldwin2020intro} also distinguishes between ``practical'' (the impossibility to shop) and ``psychological'' demand shocks (the wait-and see attitude adopted by consumers facing strong uncertainty.) We see three possibilities: (i) expenses in a specific good or service are just delayed but will take place later, for instance if households do not go to the restaurant this quarter, but go twice as much as they would normally during the next quarter; (ii) expenses are not incurred this quarter, but will come back to their normal level after the crisis, meaning that restaurants will have a one-quarter loss of sales; and (iii) expenses decrease to a permanently lower level, as household change their preferences in view of the `new normal'. Appendix \ref{appendix:demand} reproduces the scenario adopted by \citet{keogh2010possible}, which distinguishes between delay and permanently lost expenses.

\paragraph{Other components of aggregate demand.}
We do not include direct shocks to investment, net exports, and net inventories.
Investment is typically very pro-cyclical and is likely to be strongly affected, with direct factors including cash-flow reductions and high uncertainty \citep{boone2020}.
The impact on trade is likely to be strong and possibly permanent \citep{baldwin2020intro}, but would affect exports and imports in a relatively similar way, so the overall effect on net exports is unclear. Finally, it is likely that due to the disruption of supply chains, inventories will be run down so the change in inventories will be negative \citep{boone2020}.

\section{Combining supply and demand shocks}
\label{sec:combining_supply_demand}

Having described both supply and demand-side shocks, we now compare the two at the industry and occupation level. 

\subsection{Industry-level supply and demand shocks.} 
\begin{figure}[H]
    \centering
\includegraphics[width = 0.85\textwidth]{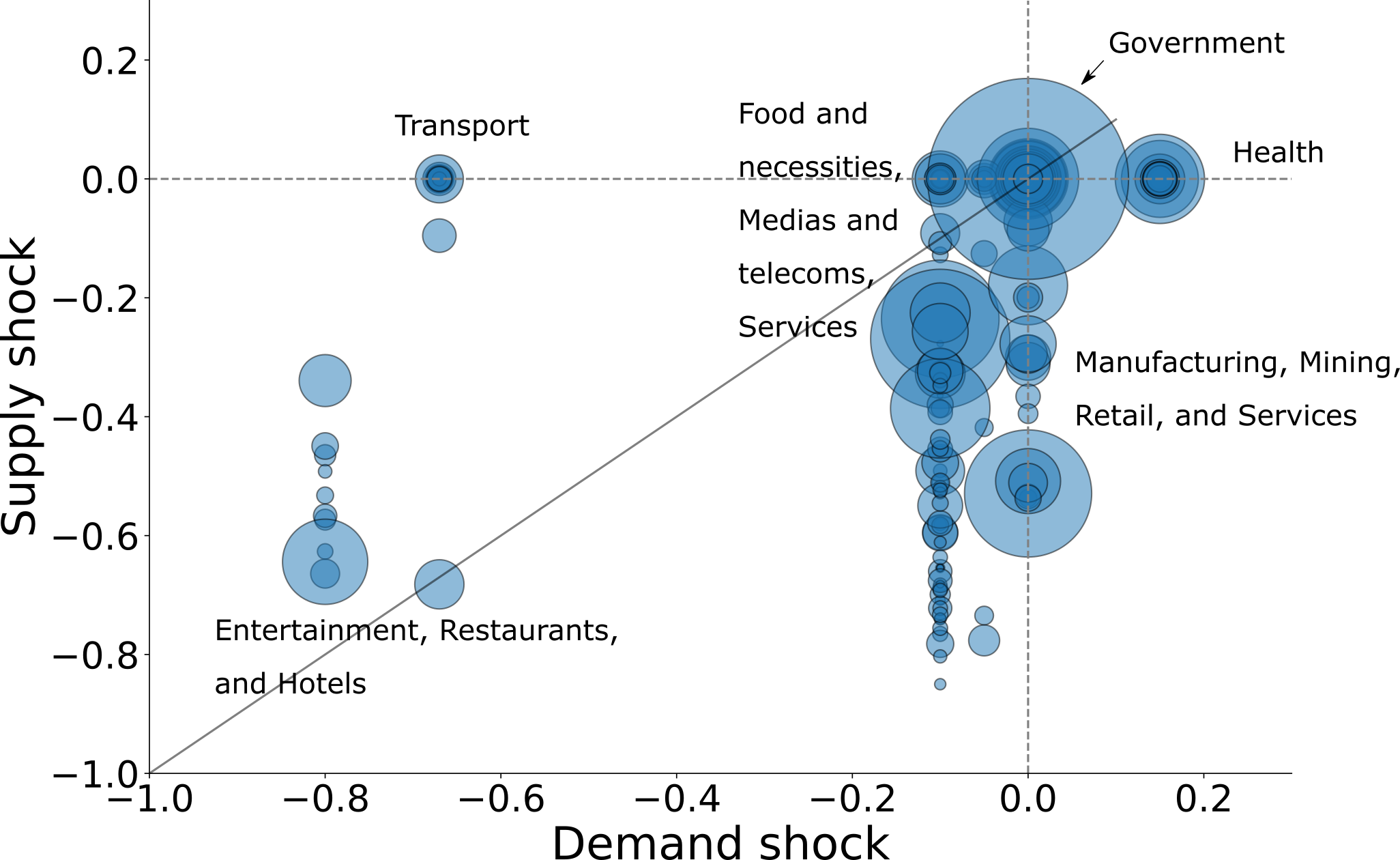}
    \caption{{\bf Supply and demand shocks for industries}.  Each circle is an industry, with radius proportional to gross output.  Many industries experience exactly the same shock, hence the superposition of some of the circles. Labels correspond to broad classifications of industries.} 
        \label{fig:supply_and_demand1}
\end{figure}

Figure \ref{fig:supply_and_demand1} plots the demand shock against the supply shock for each industry.  The radius of the circles is proportional to the gross output of the industry\footnote{
Since relevant economic variables such as total output per industry are not extensively available on the NAICS 4-digit level, we need to further aggregate the data. We derive industry-specific total output and value added for the year 2018 from the BLS input-output accounts, allowing us to distinguish 169 private industries for we which we can also match the relevant occupation data.
The data can be downloaded from \url{https://www.bls.gov/emp/data/input-output-matrix.htm}.
}.

Essential industries have no supply shock and so lie on the horizontal `0' line. Of these industries, sectors such as Utilities and Government experience no demand shock either, since immediate demand for their output is assumed to remain the same. Health experiences an increase in demand and consequently lies below the identity line. Transport on the other hand experiences a reduction in demand and lies well above the identity line. This reflects the current situation, where trains and buses are running because they are deemed essential, but they are mostly empty\footnote{
Some transport companies such as Transport for London have reduced their traffic, but this is a second-order impact, not a first-order shock. This decrease in output resulting from reduced demand should be modelled as an endogenous effect, not a first-order shock.
}. 
Non-essential industries such as Entertainment, Restaurants and Hotels, experience both a demand reduction (due to consumers seeking to avoid infection) and a supply reduction (as many workers are unable to perform their activities at home). Since the demand shock is bigger than the supply shock they lie above the identity line. Other non-essential industries such as manufacturing, mining and retail have supply shocks that are larger than their demand shocks and consequently lie below the identity line.

\subsection{Occupation-level supply and demand shocks.} 

\begin{figure}[H]
    \centering
\includegraphics[width = 1.0\textwidth]{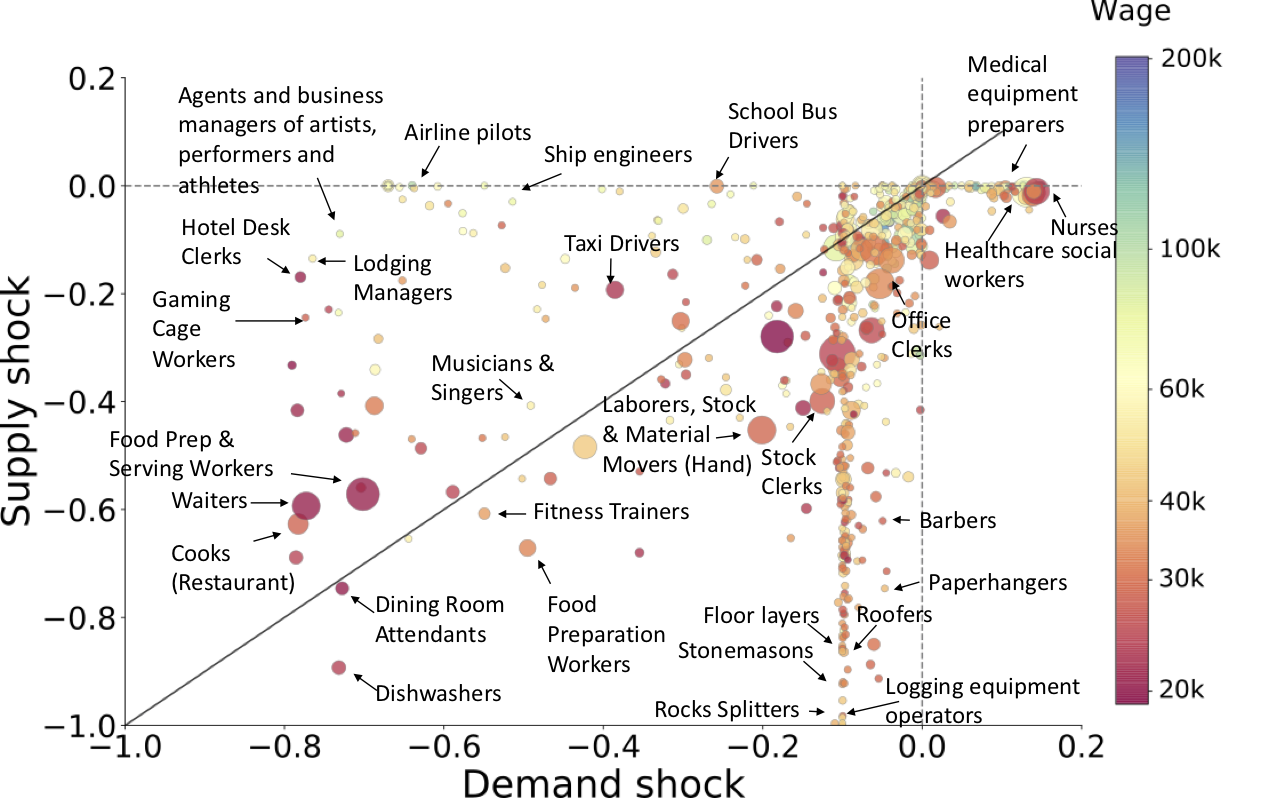}
    \caption{\textbf{Supply and demand shocks for occupations.} Each circle is an occupation with radius proportional to employment.  Circles are color coded by the log median wage of the occupation. The correlation between wages and demand shocks is 0.26 (p-value $= 2.4 \times 10^{-13}$) and between wages and supply shocks is 0.40 (p-value $= 3.9 \times 10^{-29}$).}
        \label{fig:occ_supply_demand1}
\end{figure}

In Figure \ref{fig:occ_supply_demand1} we show the supply and demand shocks for occupations rather than industries. For each occupation this comparison indicates whether it faces a risk of unemployment due the lack of demand or a lack of supply in its industry.

Several health-related occupations such as Nurses, Medical Equipment Preparers and Healthcare Social Workers are employed in industries experiencing increased demand. Occupations such as Airline Pilots, Lodging Managers and Hotel Desk Clerks face relatively mild supply shocks and strong demand shocks (as consumers reduce their demand for travel and hotel accommodation) and consequently lie above the identity line. Other occupations such as Stonemasons, Rock Splitters, Roofers and Floor Layers face a much stronger supply shock as it is very difficult for these workers to perform their job at home. Finally occupations such as Cooks, Dishwashers and Waiters suffer both adverse demand shocks (since demand for restaurants is reduced) and supply shocks (since they cannot work from home and tend not to work in essential industries). 

For the majority of occupations the supply shock is larger than the demand shock. This is not surprising given that we only consider immediate shocks and no feedback-loops in the economy. We expect that once second-order effects are considered the demand shocks are likely to be much larger.

\subsection{Aggregate shocks}

We now aggregate shocks to obtain estimates for the whole economy. We assume that, in a given industry, the total shock will be the worse of the supply or demand shocks. The shock on occupations depends on the prevalence of each occupation in each industry (see Appendix \ref{apx:total_shocks} for details). We then aggregate shocks in three different ways.

First we estimate the decline in employment by weighting occupation-level shocks by the number of workers in each occupation. Second, we estimate the decline in total wages paid by weighting occupation-level shocks by the share of occupations in the total wage bill. Finally, we estimate the decline in GDP by weighting industry level shocks by the share of industries in GDP.

Table \ref{tab:shock} shows the results. In all cases, by definition, the total shock is larger than both the supply and demand shock, but smaller than the sum. Overall, the supply shock appears to contribute more to the total shock than does the demand shock.

The wage shock is around 17\% and is lower than the employment shock (24\%). This makes sense, and reflects a fact already well acknowledged in the literature \citep{ONS2020,adams2020} that occupations that are most affected tend to have lower wages. We discuss this more below. 

The shock on value added (22\%) is higher than the shock to the wage bill. This also makes sense; because value added is the sum of the wage bill and payment to capital, this suggests that capital intensive industries are hit harder.

\begin{table}[H]
    \centering
    \begin{tabular}{|l|r|r|r|}
    \hline
    Aggregate Shock & Employment & Wages  & Value Added \\
    \hline
      Supply  &  -21 &-15  & -18 \\ 
      Demand & -13 & -8  & -8 \\ 
        \hline      
        Total &  -24 & -17 & -22 \\
        \hline
    \end{tabular}
    \caption{\textbf{Aggregate shocks to employment, wages and value added}. The size of each shock is shown as a percentage of the pre-pandemic value. Demand shocks include positive values for the health sector. The total shock at the industry level is the minimum of the supply and demand shock, see Appendix \ref{apx:total_shocks}. Note that these are only first order shocks (not  total impact), and instantaneous values (not annualized).}
    \label{tab:shock}
\end{table}

For industries and occupations in the health sector, which have experienced an increase in demand, there is no corresponding increase in supply. 
Table \ref{tab:shock_health} in Appendix \ref{apx:health_positive} provides the same estimates as Table \ref{tab:shock}, but now assuming that the increased demand for health will be matched by increased supply. This corresponds to a scenario where the healthcare sector would be immediately able to hire as many workers as necessary and pay them at the normal rate. This assumption does not, however, make a significant difference to the aggregate total shock. In other words, the increase in activity in the health sector is unlikely to be large enough to compensate significantly for the losses from other sectors.

\subsection{Shocks by wage level}
\label{section:shock_by_wage}
To understand how the pandemic has affected workers of different income levels differently, we present results for each wage quartile. The results are in Table \ref{tab:shock}, columns $q_1\dots q_4$.\footnote{
As before, Table \ref{tab:shock_health} in Appendix \ref{apx:health_positive} gives the results assuming positive total shocks for the health sector, but shows that it makes very little difference.
}
In Table \ref{tab:shock_wages} we show employment shocks by wage quartile. This table shows that workers whose wages are in the lowest quartile (lowest 25\%) will bear much higher relative losses than workers whose wages are in the highest quartile. Our results confirm the survey evidence reported by the \citet{ONS2020} and \citet{adams2020}, showing that low-wage workers are more strongly affected by the COVID crisis in terms of lost employment and lost income. Furthermore, Table \ref{tab:shock_wages} shows how the total loss of wages in the economy in split amongst the different quartiles. Even though the lowest quartile have lower salaries, the shock is so high that they bear the highest share of the total loss.

\begin{table}[H]
    \centering
    \begin{tabular}{|l|rrrr|r|}
    \hline
     & $q_1$ & $q_2$ & $q_3$ & $q_4$ &   Aggregate \\ 
        \hline
        Percentage change in employment &  -42 & -24 & -21 & -7 & -24  \\
        Share of total lost wages (\%) &  30 & 23 & 29 & 18 & -17  \\
        \hline
    \end{tabular}
    \caption{\textbf{Total Wages or employment shocks by wage quartile}.
    We divide workers into wage quartiles based on the average wage of their occupation ($q_1$ corresponds to the 25\% least paid workers). The first row is the number of workers who are vulnerable due to the shock in each quartile divided by the total who are vulnerable.  Similarly, the second row is the fraction of whole economy total wages loss that would be lost by vulnerable workers in each quartile.  The last column gives the aggregate shocks from Table 4. }
    \label{tab:shock_wages}
\end{table}

\begin{figure}[H]
    \centering
\includegraphics[width = 1\textwidth]{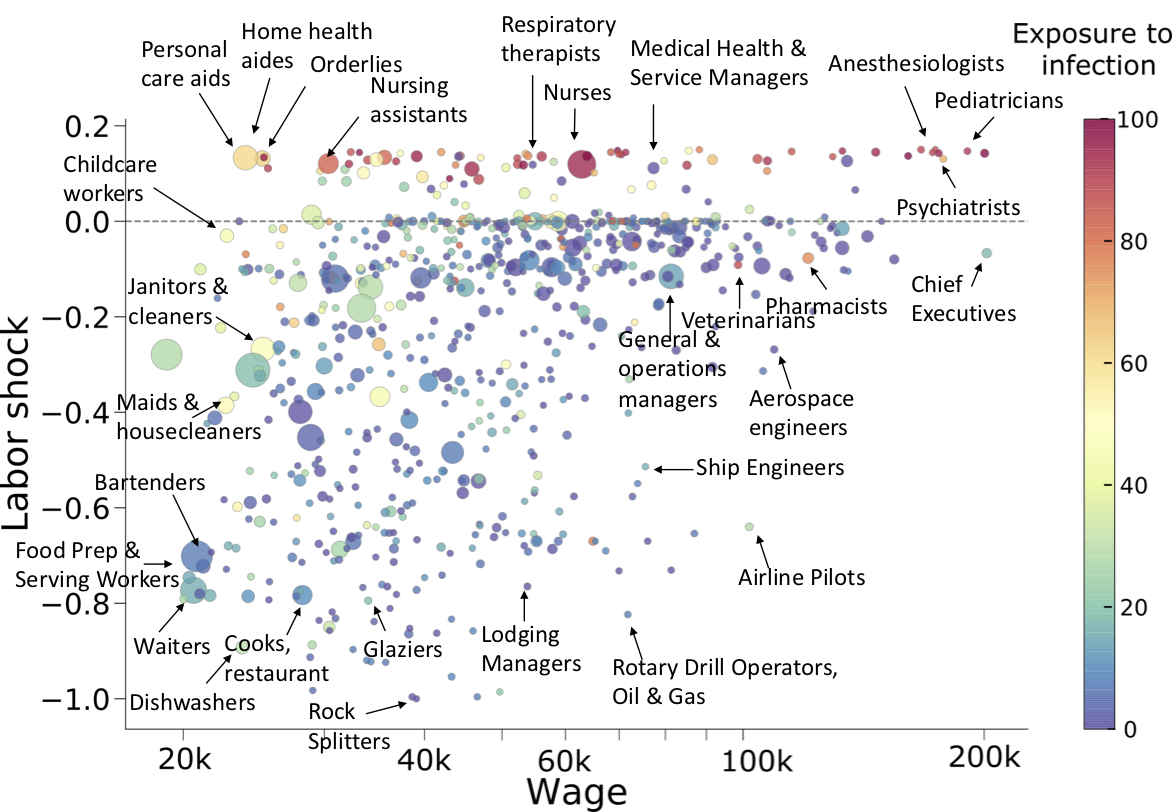}
    \caption{\textbf{Labor shock vs. median wage for different occupations.} We color occupations by their exposure to disease and infection.  There is a $0.40$ correlation between wages and the labor shock (p-value $= 4.4\times 10^{-40}$). Note the striking lack of high wage occupations with large labor demand shocks.}
        \label{fig:occ_supply_demand2}
\end{figure}

Next we estimate labor shocks at the occupation level.  We define the labor shocks as the declines in employment due to the total shocks in the industries associated with each occupation. We use Eq. (\ref{eq:occ_totalshock_v2}) (Appendix \ref{apx:health_positive}) to compute the labor shocks, which allows for positive shocks in healthcare workers, to suggest an interpretation in terms of a change in labor demand. Figure \ref{fig:occ_supply_demand2} plots the relationship between labor shocks and median wage. A strong positive correlation (Pearson $\rho = 0.40$, $\text{p-value} = 4.4\times 10^{-30})$ is clearly evident,  with almost no high wage occupations facing a serious shock.

We have also colored occupations by their exposure to disease and infection using an index developed by O*NET\footnote{
\url{https://www.onetonline.org/find/descriptor/result/4.C.2.c.1.b}
} (for brevity we refer to this index as ``exposure to infection’’). As most occupations facing a positive labor shock relate to healthcare\footnote{
Our demand shocks do not have an increase in retail, but in the UK supermarkets have been trying to hire several tens of thousands workers (Source: BBC, 21 March, \url{https://www.bbc.co.uk/news/business-51976075}). \citet{baker2020how} documents stock piling behavior in the US.},
it is not surprising to see that they have a much higher risk of being exposed to disease and infection. However, other occupations such as janitors, cleaners, maids and childcare workers also face higher risk of infection. Appendix \ref{sec:occupations_at_risk} explores the relationship between exposure to infection and wage in more detail.

\section{Conclusion}\label{sec:discussion}

This paper has sought to provide quantitative predictions for the U.S. economy of the supply and demand shocks associated with the COVID-19 pandemic. To characterize supply shocks, we developed a Remote Labor Index to estimate the extent to which workers can perform activities associated with their occupation at home and identified which industries are classified as essential vs. non-essential. We also reported plausible estimates of the demand shocks, in an attempt to acknowledge that some industries will have an immediate reduction in output due to a shortfall in demand, rather due to an impossibility to work.  We would like to emphasize that these are {\it predictions}, not measurements:  The estimates of the demand shocks were made in 2006, and the RLI and the list of non-essential industries contain no pandemic-specific information, and could have been made at any time.
Putting these predictions together, we estimate that the first order aggregate shock to the economy represents a reduction of roughly a quarter of the economy.

This is the first study seeking to compare supply-side shocks with corresponding demand-side shocks at the occupation and industry level. At the time of writing, the most relevant demand-side estimates available are highly admittedly `rough' and only available for very aggregate (2-digit) industries. Yet, this suggests that sectors such as Transport are more likely to have output constrained by demand-side shocks, while sectors relating to manufacturing, mining and services are more likely to be constrained by supply-side shocks. Entertainment, restaurants and tourism face both very large supply and demand constraints, with demand shocks dominating in our estimates. By quantifying supply and demand shocks by industry, our paper speaks to the debate on the possibility of inflation after the crisis. \citet{goodhart2020} argue that the lockdown causes a massive supply shock that will lead to inflation when demand comes back after the crisis. But as \citet{miles2020} note, in many sectors it is not obvious that demand will come back immediately after the crisis, and if a gradual reopening of the economy takes place, it may be that supply and demand rise slowly together. However, our paper is the first to raise the fact that because supply and demand shocks are so different by sectors, even a gradual reopening may leave important supply-demand imbalances within industries. Such mismatches could consequently lead to an unusual level of heterogeneity in the inflation for different goods.

When considering total shocks at the occupation level, we find that high-wage occupations are relatively immune from both supply and demand-side shocks, while many low-wage occupations are much more economically vulnerable to both. Interestingly, low-wage occupations that are \textit{not} vulnerable to supply and/or demand-side shocks are nonetheless at higher risk of being exposed to coronavirus (see color code in Figure \ref{fig:occ_supply_demand2}). Such findings suggest that the COVID-19 pandemic is likely to exacerbate income inequality in what is already a highly unequal society.

For policymakers there are three key implications from this study.  First, the magnitude of the shocks being experienced by the U.S. economy is very large, with around a quarter of the economy not functioning. 
As Table 4 shows, even including positive shocks, our estimates of the potential impacts are a drop in employment of 24\%, a decline in wages of 17\%, and loss in value added of 22\%.  Bearing in mind the caveats about shocks vs. total impacts, the potential impacts are a multiple of what was experienced during the global financial crisis (e.g. where employment dropped 3.28 percentage points)\footnote{
Employment Rate, aged 15-64, all persons for the US (FRED LREM64TTUSM156N) fell from 71.51 in December 2007
to 68.23 in June 2009, the employment peak to trough during the dates of recession as defined by the NBER.
}
and comparable only to the Great Depression (e.g. where employment dropped 21.7\% 1929-32 \citep[Table 2]{Wallis1989}).
Second, as the largest shocks are from the supply-side, strategies for returning people to work as quickly as possible without endangering public health must be a priority.  Virus mitigation and containment are clearly essential first steps, but strategies such as widespread antibody testing to identify people who are safe to return to work, and rapid testing, tracing, and isolation to minimize future lock-downs, will also be vital until if and when a vaccine is available.  Furthermore, aggressive fiscal and monetary policies to minimize first-order shocks cascading into second-order shocks are essential, in particular policies to keep workers in employment and maintain incomes (e.g. the “paycheck protection” schemes announced by several countries), as well as policies to preserve business and financial solvency.  Third, and finally, the inequalities highlighted by this study will also require policy responses.  Again, higher income knowledge and service workers will likely see relatively little impact, while lower income workers will bear the brunt of the employment, income, and health impacts.  In order to ensure that burdens from the crisis are shared as fairly as possible, assistance should be targeted at those most effected, while taxes to support such programs be drawn primarily from those least effected.

To reiterate an important point, our predictions of the shocks are \emph{not}  estimates of the overall impact of the COVID-19 on the economy, but are rather estimates of the first-order shocks. Overall impacts can be very different form first order shocks for several reasons: First, shocks to a particular sector propagate and may be amplified as each industry faces a shock and reduces its demand for intermediate goods from other industries.
Second, industries with decreased output will stop paying wages of furloughed workers, thereby reducing income and demand.  Third, the few industries facing higher demand will increase supply, if they can overcome labor mobility frictions \citep{del2019automation}. We make our predictions of the shocks available here so that other researchers can improve upon them and use them in their own models\footnote{Our data repository is at \url{https://zenodo.org/record/3751068}, where we will post any update.}. We intend to update and use these shocks ourselves in our models in the near future.

We have made a number of strong assumptions and used data from different sources. To recapitulate, we assume that the production function for an industry is linear and that it does not depend on the composition of occupations who are still able to work; we neglect absenteeism due to mortality and morbidity, as well as loss of productivity due to school closures (though we have argued these effects are small -- see Appendix \ref{appendix:epidemiology}).  We have constructed our remote labor index based on a subjective rating of work activities and we assumed that all work activities are equally important and they are additive.  We have also applied a rating of essential industries for Italy to the U.S.  Nonetheless, we believe that the analysis here provides a useful starting point for macroeconomic models attempting to measure the impact of the COVID-19 pandemic on the economy.

As new data becomes available we will be able to test whether our predictions are correct and improve our shock estimate across industries and occupations. Several countries have already started to release survey data. New measurements about the ability to perform work remotely in different occupations are also becoming available. New York and Pennsylvania have released a list of industries that are considered essential\footnote{\url{https://esd.ny.gov/guidance-executive-order-2026}} (though this is not currently associated with any industrial classification such as NACE or NAICS). As new data becomes available for the mitigation measures different states and countries are taking, we can also refine our analysis to account for different government actions. Thus we hope that the usefulness of methodology we have presented here goes beyond the immediate application, and will provide a useful framework for predicting economic shocks as the pandemic develops.

\small
\bibliographystyle{agsm}
\bibliography{tech_ref}
\FloatBarrier
\normalsize
\appendix
\section*{Appendix}
\label{sec:Appendix}

\section{Derivation of total shocks} 
\label{apx:total_shocks}

\subsection{Derivation of supply shocks} 
\label{apx:supply_shocks}


As discussed in the main text, we estimate the supply shock by computing an estimate of the share of work that will not be performed, which we compute by estimating the share of work that is not in an essential industry and that cannot be performed from home. We had to use several concordance tables, and make a number of assumptions, which we describe in details here.

\begin{figure}[!ht]
    \centering
\includegraphics[page=1, width = \textwidth]{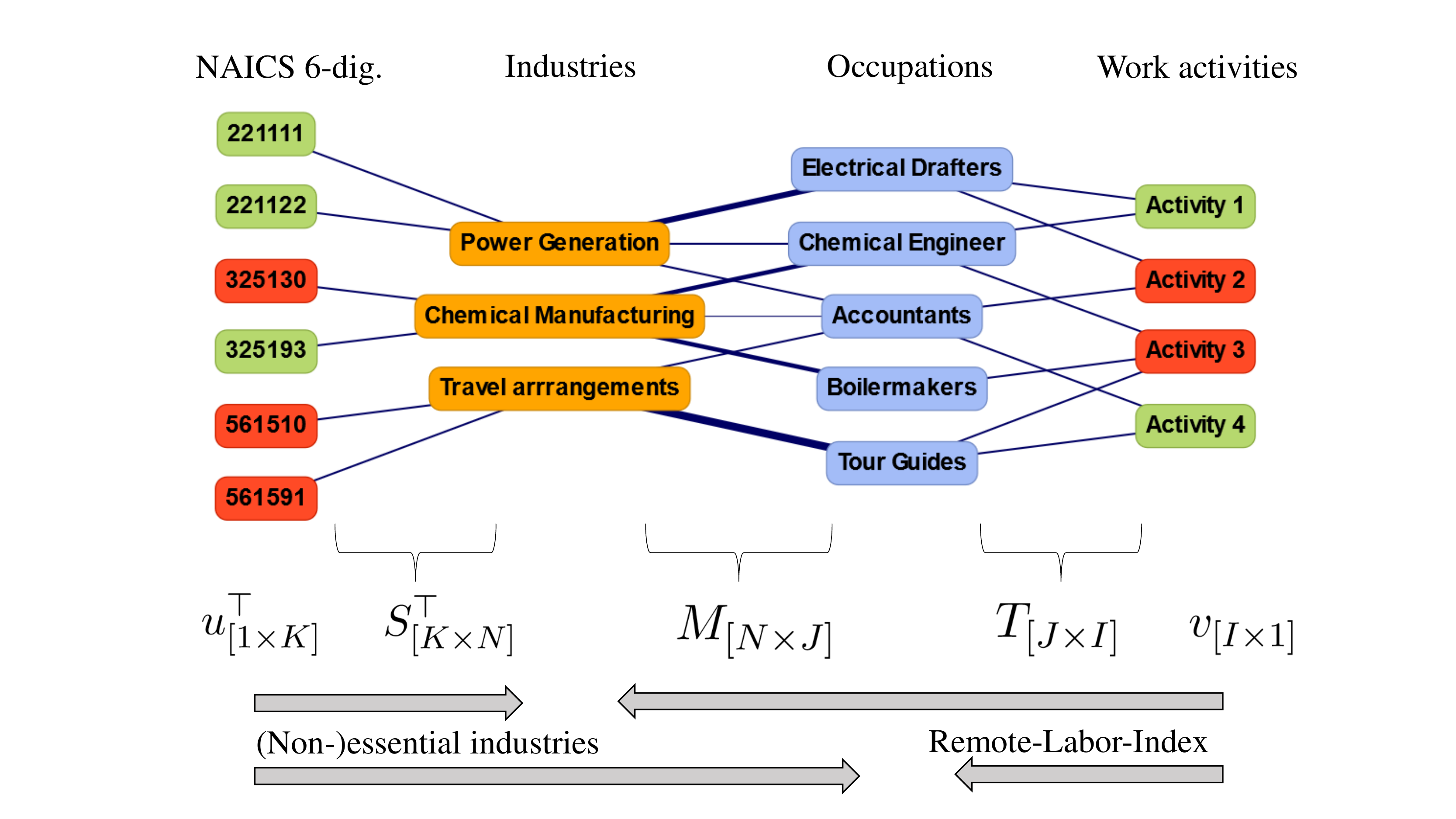}
    \caption{The same schematic network representation of supply side shocks as in the main text, but now also including mathematical notation. 
    The $K$-dimensional vector $u$ below the NAICS 6-dig. (left nodes) encodes essential industries with binary elements. 
    This set of nodes is connected to relevant industry categories by concordance tables (incidence matrix $S$).
    Matrix $M$ connects the $N$ industry categories with $J$ occupations where an element represents the corresponding employment number. The ability to perform work activities (right nodes) from home is represented in vector $v$, also by binary elements.
    We use occupation-activity mappings provided from O*NET, represented as incidence matrix $T$.
    The grey arrows show the direction of shocks to industries and employment. The shock originating from the list of essential industries is mapped directly onto the broader industry categories, before it can be computed for occupations. Conversely, the remote-labor-index is first mapped onto occupations and then projected onto industries. 
    }
        \label{fig:supply_shocks}
\end{figure}


Figure \ref{fig:supply_shocks} illustrates our method. There are four sets of nodes which are connected by three bipartite networks. 
The first set of nodes are the 6-digit NAICS industries which are classified to be essential or non-essential. This information is encoded in the $K$-dimensional column vector $u$ which element $u_k=1$ if NAICS 6-digit industry $k$ is essential and 0 otherwise.
Second, there are $N$ different industry categories on which our economic analysis is based. The 6-digit NAICS codes are connected to these industries by the incidence matrix (concordance table) $S$.
The third set of nodes are the $J$ occupations obtained from the BLS and O*NET data. The weighted incidence matrix $M$ couples industries with occupations where the element $M_{nj}$ denotes the number of people in occupation $j$ being employed in industry $n$. 
Fourth, we also have a list of $I$ work activities. Each activity was rated whether it can be performed from home. If activity $i$ can be done from home, the $i^{\text{th}}$ element of the vector $v$ is equal 1, and otherwise it is equal to 0.
The incidence matrix $T$ denotes whether an occupation is associated with any given work activity, i.e $T_{ji}=1$ if activity $i$ is relevant for occupation $j$. 

The analysis presented here is based on $I=332$ unique work activities, $J=740$ occupations and $K=1,057$ 6-digit NAICS industries. When relating to industry-specific results we use the BLS industry categories of the input-output accounts, leaving us with $N=169$ industries for which we have reliable data on value added, total output and other key statistics.
Employment, occupation and wage statistics are available on a more fine-grained 4-digit NAICS level. We therefore use these $N=277$ industries for deriving labor-specific results.

\subsection{Industry-specific shocks}
We can use this simple framework for deriving the supply shocks to industries.

\paragraph{(Non-)essential industries.}
To estimate the extent to which an industry category is affected by a shutdown of non-essential economic activities, we measure the fraction of its 6-digit NAICS sub-industries which are classified as non-essential. In mathematical terms, the essential-score for every industry is therefore a weighted sum which can be written compactly in matrix notation as
\begin{equation} \label{eq:industry_essential}
    e = \Tilde{S} u,
\end{equation}
where $\Tilde{S}$ is the row-normalized version of matrix $S$ with elements $\Tilde{S}_{nk} =  S_{nk}/\sum_{h} S_{nh}$.

Note that this assumes that the fine-grained NAICS codes contribute uniformly to the more aggregate industry categories. Although this assumption might be violated in several cases, in absence of further information, we use this assumption throughout the text.


\paragraph{Industry Remote-Labor-Index.}
We can similarly estimate the extent to which the production of occupations or industries can take place by working from home. Since work activities are linked to occupations, but not directly to industries, we need to take two weighted averages to obtain the industry-specific RLI.

For each occupation we first measure the fraction of work activities that can be done from home. We interpret this as the share of work of an occupation that can be performed from home, or `occupation-level RLI'. This interpretation makes two assumptions: (i) that every work activity contributes equally to an occupation, which is our best guess since we do not have better data, and (ii) that if $z$\% of activities cannot be done from home, the other $1-z\%$ of activities can still be carried out and and are as productive as before.

For each industry $i$ we then take a weighted average of the occupation-level RLIs, where the weights are the shares of workers employed in each occupation and in industry $i$. Let $\Tilde{T}$ denote the row-normalized version of matrix $T$, i.e $\Tilde{T}_{ji} = T_{ji}/\sum_{h} T_{jh}$ and similarly let the element of matrix $\Tilde{M}$ be $\Tilde{M}_{nj} = M_{nj}/\sum_{h} M_{nh}$. Then the industry-specific remote-labor-index is given by the vector
\begin{equation} \label{eq:industry_rli}
    r = \Tilde{M} \Tilde{T} v.
\end{equation}

We interpret the remote-labor-index for an industry, $r_n$, as the fraction of work in an industry $n$ that can be performed from home. As for assumption (ii) above for the occupation-level RLI, this assumes that if $z$\% of the work of occupations cannot be done, the other $1-z\%$ of work can still be carried out.

\paragraph{Immediate industry supply shock.}
To derive industry supply shocks from the scores above, we need to take into account that industries might be exposed to both effects at the same time, but with different magnitudes. For example, consider the illustrative case of \textit{Chemical Manufacturing} in Figure \ref{fig:supply_shocks}. Half of the industry is non-essential (red node `325130') and could therefore be directly affected by an economic shutdown. But different occupations can be found in this industry that are affected heterogeneously. In this simple example, \textit{Chemical Manufacturing} draws heavily on Boilermakers who have only work activities that cannot be done from home. On the other hand, this industry also has a tiny share of accountants and a larger share of Chemical Engineers who are able to do half of their work activities from home.

As stated above, the essential score $e_n$ and the RLI $r_n$ can be interpreted as shares of industry-specific work which can be performed, either thanks to being essential or thanks to being adequately done from home. To compute the share of industry-specific work that can performed due to either effect, we interpret shares as probabilities and assume independence,
\begin{equation} \label{eq:industry_supplyshock}
 \text{ISS}_n =  -\left(1 - e_n\right) \left(  1 - r_n  \right),
\end{equation}
where ISS stands for `Industry Supply Shock'.  We have multiplied the probability by minus one to obtain negative shocks.
Although independence is a strong assumption, we have no reason to believe that the work that can be done from home is more or less likely to be judged essential. The empirical correlation coefficient of $e$ and $r$ is 0.03 and is far from being significant (p-value of 0.7). Under the linear assumptions about combining labor we are making here this indicates independence.  

When applying these Industry Supply Shocks to value added, we make the implicit assumption that 
a $z$\% decrease in labor will cause a $z$\% decrease in value added.

\subsection{Occupation-specific shocks.}
We now describe how we compute shocks for specific occupations, rather than specific industries.

\paragraph{Occupations in (non-)essential industries.}
Occupations are mapped to industries through the weighted incidence matrix $M$, where an element denotes the number of jobs per occupation and industry. The column-normalized matrix $M^*$ with elements $M^*_{nj} = M_{nj}/\sum_h M_{hj}$ denotes the share of an occupation carried out in a particular industry\footnote{Note that we column-normalize $M$ to map from industries to occupations and row-normalize when mapping from occupations to industries.}. The essential-score for occupations is taken as weighted average of the essential score for industries (computed in Eq. \ref{eq:industry_essential}),
\begin{equation} \label{eq:occupation_essential}
    x = M^{* \top} \Tilde{S} u =  M^{* \top} e.
\end{equation}

\paragraph{Occupation Remote Labor Index.}
As already indicated in the derivation of the industry-specific RLI, $r$, in Eq. (\ref{eq:industry_rli}), the occupation-specific RLI, $y$, is a weighted average of all the corresponding work activities that can be done from home. Formally, the occupation-based RLI is given by
\begin{equation} \label{eq:occupation_rli}
    y = \Tilde{T} v.
\end{equation}

\paragraph{Total supply-driven occupation shock.}
Following the same procedure as in Eq. \ref{eq:industry_supplyshock}, we can get the total immediate shock on occupations from the economy's supply side\footnote{To be clear, this is a product market \emph{supply-side} shock, but this translates into a reduction in labor \emph{demand} in each occupations.}.
The combined immediate shock to occupations is then given as 
\begin{equation} \label{eq:occupation_supplyshock}
 \text{OSS}_j =  -\left(1 - x_j\right) \left(  1 - y_j  \right).
\end{equation}

Here, the correlation between RLI and the essential-score is larger, $\rho (x,y)=0.30$ (p-value $= 8.3 \times 10^{-17}$), and significant which can also be seen from Figure \ref{fig:occ_ind_broad_panel}. It should therefore be noted that the labor-specific results are expected to be more sensitive with respect to the independence assumption, as it is the case for industry-related results.

\subsection{Derivation of demand shocks}
Since we have demand shocks only on the 2-digit NAICS level, disaggregating them into the more fine-grained relevant industry categories is straightforward when assuming that the demand shock holds equally for all sub-industries. We let the industry demand shock in percentages for industry $n$ be $-\text{IDS}_n$.

To map the demand shocks onto occupations, we can invoke the same matrix algebra as above. The occupation-specific shock originating from the economy's demand side is then given by the projection
\begin{equation} 
    \text{ODS} = M^{* \top} \text{IDS}.
\end{equation}

\subsection{Total immediate (first-order) shocks}

We now combine supply- and demand-driven shocks to total immediate shocks for occupations and industries. 

Let us turn to industries first. As discussed in more depth in the main text, the shock experienced is likely to be the worse of the two (supply and demand) shocks. Since we have expressed shocks as negative if they lead to decrease in output, in more mathematical terms, the \textit{industry total shock} then is
\begin{equation}
\text{ITS}_n =
    \min(\text{ISS}_n, \text{IDS}_n).
    \label{eq:ITSn}
\end{equation}
and the \textit{occupation total shock} is
\begin{equation}
\text{OTS}_j =
    \min(\text{OSS}_j, \text{ODS}_j).
    \label{eq:OTSj}
\end{equation}

Note that under these assumptions, the health sector will not experience a positive shock. Our estimates of the shocks suggest that there is an increase in demand, but we have not described positive supply shocks for the health sector. As a result, industries and occupations in the health sector will have a non-positive total shock. 

Of course, the actual production of health care is likely to go up. But this is the result of an adjustment to an increase demand, rather than a first-order shock, and is therefore not modelled here. Nevertheless, see Appendix \ref{apx:health_positive} for an alternative.

\subsection{Aggregate total shocks}

To provide economy-wide estimate of the shocks, we aggregate industry- or occupation-level shocks. We do this using different sets of weights. 

First of all, consider the interpretation that our shocks at the occupation-level represent the share of work that will not be performed. If we assume that if $z\%$ of the work cannot be done, $z\%$ of the workers will become unemployed, we can weight the occupation shocks by the share of employment in each occupation. Using the vector $L$ to denote the share of employed workers that are employed by occupation $j$, we have 
\begin{equation}
\text{Employment total shock} = \text{OTS}^T L
\end{equation}
The Employment supply (demand) shock is computed similarly but using OSS (ODS) instead of OTS. 

Instead of computing how many workers may lose their job, we can compute by how much paid wages will decrease. For each occupation, we compute the total wage bill by multiplying the number of workers by the average wage. We then create a vector $w$ where $w_j$ is the share of occupation $j$ in the total wage bill. Then, 
\begin{equation}
\text{Wage total shock} = \text{OTS}^T w,
\end{equation}
and similarly for the OSS and ODS. Note that we omit three occupations for which we do not have wages (but had employment).

Finally, to get an estimate of the loss of GDP, we can aggregate shocks by industry, weighting by the share of an industry in GDP. Denoting by $Y$ the vector where $Y_n$ is the VA of industry $n$ divided by GDP\footnote{Our estimate of GDP is the sum of VA of industries in our sample.},
\begin{equation}
\text{Value added total shock} = \text{ITS}^T Y,
\end{equation}
and similarly for the industry supply and demand shocks (ISS and IDS). Note that we could have used shares of gross output and compute a shock to gross output rather to GDP.

\subsection{Aggregate total shocks with growth of the Health sector}
\label{apx:health_positive}
Here we make a different assumption about how to construct the total shock for occupations and industries. For industries, we assume that if they experience a positive demand shock, the industries are able to increase their supply to meet the new demand. Instead of Eqs. (\ref{eq:ITSn}) we use
\begin{equation}
    \text{ITS}^h_n =
    \begin{cases}
      \text{ITS}_n, & \text{if}\ \text{IDS}_n \le 0 \\
      \text{IDS}_n, & \text{if}\ \text{IDS}_n > 0. \\
    \end{cases}
    \label{eq:ind_totalshock_v2}
\end{equation}
Since occupations are employed by different industries, the total shock to an occupation can be influenced by positive demand shocks from the healthcare sector and negative demand shocks from non-essential industries. 
In Eq. (\ref{eq:OTSj}) we consider that occupations only experience the negative shocks. An alternative is to consider both the negative shock caused by non-essential industries and the positive shock caused by the health industries. This gives 
\begin{equation}
    \text{OTS}^h_j =
    \begin{cases}
      \text{OTS}_j, & \text{if}\ \text{ODS}_j \le 0 \\
      \text{ODS}_j + \text{OSS}_j, & \text{if}\ \text{ODS}_j>0. \\
    \end{cases}
    \label{eq:occ_totalshock_v2}
\end{equation}
In Section \ref{section:shock_by_wage}, specifically 
Figure \ref{fig:occ_supply_demand2} we use this convention for the y-axis, the Labor Shock. Using Eq. (\ref{eq:occ_totalshock_v2}) allows us to observe how health related occupations experience a positive shock. 

In Table \ref{tab:shock_health} we show the aggregate total shocks when using Eqs.(\ref{eq:ind_totalshock_v2}) and (\ref{eq:occ_totalshock_v2}). There is very little difference with the results in the main text. 
The Health sector and its increase in demand are not large enough to make a big difference to aggregate results.

\begin{table}[H]
    \centering
    \begin{tabular}{|l|rrrrr|rr|}
    \hline
     & \multicolumn{5}{c|}{Employment} &  Wages & Value Added \\
    Shock & Aggregate & $q_1$ & $q_2$ & $q_3$ & $q_4$ & Aggregate  & Aggregate  \\ 
        \hline      Total& -22 & -41 & -22 & -20  &  -4 & -15 &  -21 \\
        \hline
    \end{tabular}
    \caption{\textbf{Main results allowing for growth in the Health sector}. The results are the same as those presented in Table \ref{tab:shock}, but assuming that in industries, when demand is positive, the total shock is equal to the demand shock}
    \label{tab:shock_health}
\end{table}

\section{Data}
\label{apx:data}

In this section we give more details about how we constructed all our variables. We stress that our goal was to produce useful results quickly and transparently, and make them available so that anyone can update and use them. We intend to improve these estimates ourselves in the future, as more information becomes available on the ability to work from home, which industries are essential, and how consumers react to the crisis by shifting their spending patterns.

\subsection{Italian list of essential industries}\label{apx:nace-naicsmap}
The Italian list of essential industries\footnote{Available at \url{http://www.governo.it/sites/new.governo.it/files/dpcm_20200322.pdf}, 22 March.} is based on the Statistical Classification of Economic Activities in the European Community, commonly referred to as NACE. The list of essential industries are listed with NACE 2-digit, 4-digit and 6-digit codes. We automatically map industries listed at the 2 or 4-digit NACE level to NAICS 6-digit industries using the crosswalk made available by the European Commission\footnote{
\url{https://ec.europa.eu/eurostat/ramon/relations/index.cfm?TargetUrl=LST\_REL\&StrLanguageCode=EN\&IntCurrentPage=11}}.
The 6-digit NACE level classification is country dependent and thus there is no official crosswalk to NAICS codes. We map the 6-digit industries by hand.

In a second step, we looked at the list of resulting list of industries and their essential score and discovered a few implausible cases, resulting from the complex mapping between the various classification systems at different levels. For instance, because Transports are essential, ``Scenic and sightseeing transportation, other'' was considered essential. In contrast, ``Death care services'' was classified as non essential. Three of us, as well as two independent colleagues with knowledge of the current situation in Italy, evaluated the list and we proceeded to editing the 4-digit NAICS Essential Scores as follows: From Non-essential to Essential: Grocery stores, Health and personal care stores, Gasoline stations, Death care services. From Essential (sometimes only partly) to non essential: Scenic and sightseeing transportation, Software publishers, Motion picture and video industries, Sound recording industries, and Other amusement and recreation industries. Finally, Federal, State and Local Government were not classified, and we classified as Essential.

\subsection{Data for occupations}\label{apx:occupation}
O*NET has work activities data for 775 occupations, out of which 765 occupations have more than five work activities. We compute the remote labor index for the 765 occupations with more than five work activities. From the May 2018 Occupational Employment Statistics (OES) estimates on the level of 4-digit NAICS (North American Industry Classification System), file \textit{nat4d\_M2018\_dl}, which is available at \url{https://www.bls.gov/oes/tables.htm} under \textit{All Data}, we find data for the number of employed workers of 807 occupations in 277 industries. These data covers 144M workers\footnote{(the US economy had 156M workers mid-2018, see \url{https://fred.stlouisfed.org/series/CE16OV})}. From the sample of 765 occupations with RLI, and from the sample of 807 occupations with employment data from the BLS, we are able to match 740 occupations, which cover 136.8M workers. Therefore, our final sample has  740 occupations and 136.8M workers.\footnote{Note that the BLS employment data we use here does not include self-employed workers (which currently accounts for about 16 million people).}

With the occupation-industry employment data and the essential score of each industry, we estimate the share of essential jobs within each occupation. 
Additionally, we have wage information for most occupations (i.e. we have median and mean wage data for 737 occupations). We computed all correlations for median wage considering all occupations we had median wage data for. For the 3 occupations for which median wage. data was missing, the color coding of occupations in Figures \ref{fig:occ_ind_broad_panel}, \ref{fig:occ_supply_demand1}, and the x-axis in Figure \ref{fig:occ_supply_demand2} corresponds to the average (across all occupations) of the median wage. We used the mean wages and the employment of occupations to define the wage quartiles of our sample. We excluded the 3 occupations for which we did not have mean wage from these calculations.

Finally, we use the O*NET data on exposure to disease and infection of occupations for the color coding in Figure \ref{fig:occ_supply_demand2}. We explain these data further in Appendix \ref{sec:occupations_at_risk}.  
In the following charts we show the distribution of the remote labor index, exposure to disease and infection, supply, demand and overall shocks across occupations.

\begin{figure}[H]
    \centering
\includegraphics[width = 0.45\textwidth]{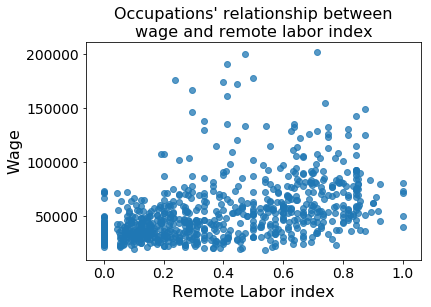}\includegraphics[width = 0.45\textwidth]{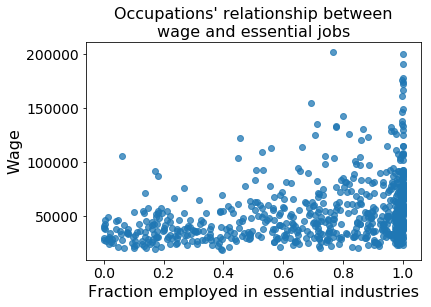}
    \caption{Left. Relationship between remote labor index and median wage. The Pearson correlation is 0.42 (p-value = $ 7.3\times 10^{-33}$). Right. Relationship between fraction of workers in essential industries and wage.  The Pearson correlation is 0.35 (p-value = $ 1.12 \times 10^{-22}$).}
        \label{fig:occ_wage_ess}
\end{figure}

\begin{figure}[H]
    \centering
\includegraphics[width = 0.45\textwidth]{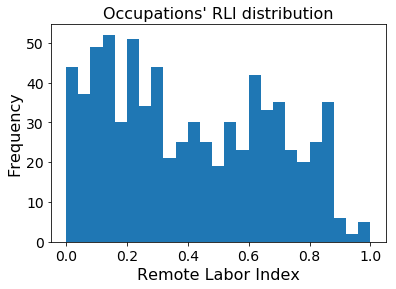}\includegraphics[width = 0.45\textwidth]{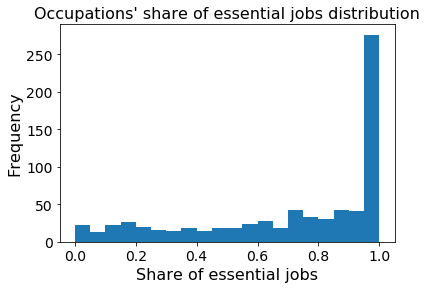}
    \caption{Left. Distribution of the remote labor index for the 740 occupations. Right. Distribution of the share of essential jobs within each of the 740 occupations.  }
        \label{fig:occ_ess_dist}
\end{figure}

\begin{figure}[H]
    \centering
\includegraphics[width = 0.45\textwidth]{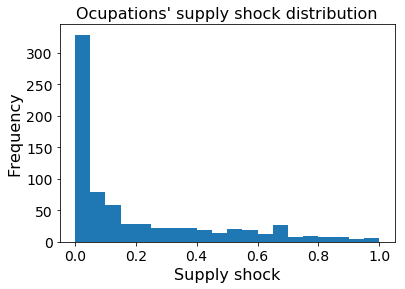}\includegraphics[width = 0.45\textwidth]{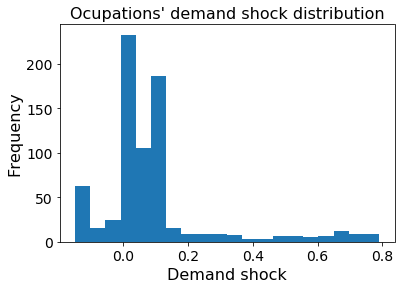}
    \caption{Left. Supply shock distribution across occupation.  Right. Demand shock distribution across occupations.}
        \label{fig:occ_demand_dist}
\end{figure}

\begin{figure}[H]
    \centering
\includegraphics[width = 0.45\textwidth]{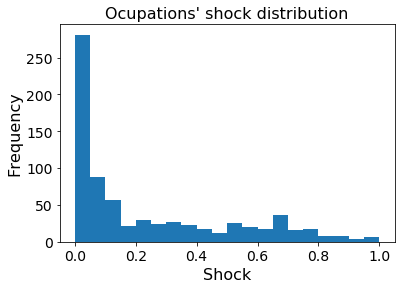}\includegraphics[width = 0.45\textwidth]{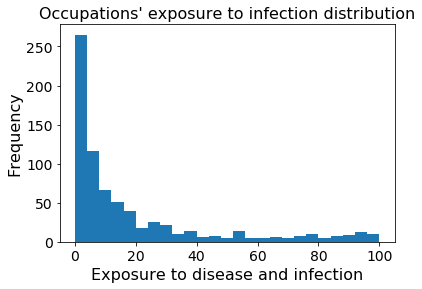}
    \caption{Left. Shock distribution for occupations.  Right. Distribution of exposure to disease.}
        \label{fig:occ_overall_inf_dist}
\end{figure}

\subsection{Data for industries}
\label{apx:industries}

\paragraph{Matching all data to BLS I-O industries.}
A key motivation of this paper is to provide relevant economic data which can be used by other researchers and policymakers to model the economic impact of the COVID-19 pandemic. We therefore bring the discussed supply and demand shock data into a format that matches directly to US input-output data.

We use the BLS 2018 input-output account, which allows us to discern 179 private sectors. Moreover, there are the additional industries \textit{Private Households}, NAICS 814 and \textit{Postal Service}, NAICS 491.
The data also contains 19 different industries relating to governmental activities. Since these industries are not classified with NAICS codes, we aggregate all governmental industries into a single node \textit{Government}, which can be interpreted as the NAICS 2-digit industry 92.
This leaves us with 182 industry categories which are a mixture of 2- to 6-digit NAICS industries.

We are able to match occupational data to 170 out of the 182 industry categories, accounting for 97\% of total value added. For this subset we compute industry-specific RLIs, essential scores and supply shocks as spelled out in Appendix \ref{apx:supply_shocks}, as well as employment-weighted infection exposures. 

Since we have demand shocks only on the 2-digit NAICS level, disaggregating them into the more fine-grained BLS input-output data is straightforward when assuming that the demand shock holds equally for all sub-industries.

In the online data repository we also report total wages and total employment per industry. We use the same OES estimates as for the occupational data, but match every industry category according to the corresponding NAICS 2- to 6-digit digit levels.

\begin{figure}[H]
    \centering
\includegraphics[width = 0.45\textwidth]{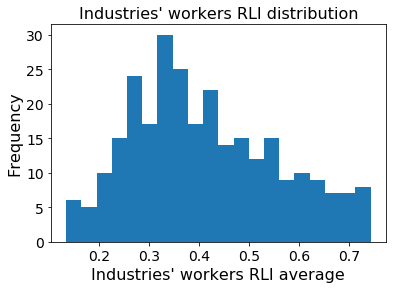}\includegraphics[width = 0.45\textwidth]{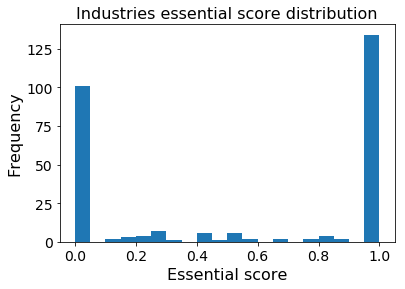}
    \caption{Left. Distribution of the remote labor index, aggregated to 169 industries. Right. Fractions of essential sub-industries per industry category.}
        \label{fig:ind_rli_ess}
\end{figure}

\begin{figure}[H]
    \centering
\includegraphics[width = 0.45\textwidth]{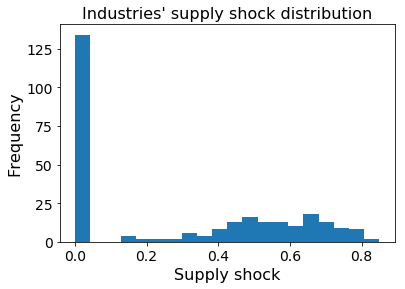}\includegraphics[width = 0.45\textwidth]{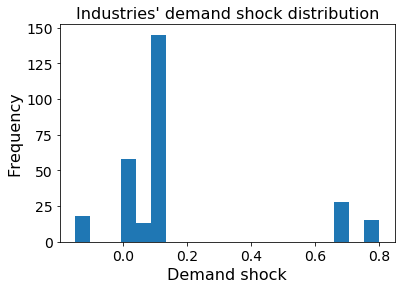}
    \caption{Left. Supply shock distribution across industries.  Right. Demand shock distribution across industries.}
        \label{fig:ind_supply_demand_dist}
\end{figure}

\begin{figure}[H]
    \centering
\includegraphics[width = 0.45\textwidth]{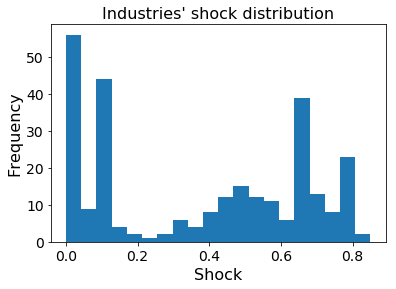}\includegraphics[width = 0.45\textwidth]{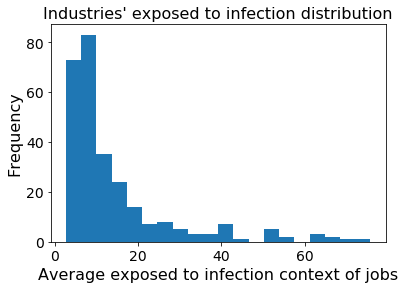}
    \caption{Left. Shock distribution across industries.  Right. Distribution of exposure to disease across industries.}
        \label{fig:ind_overall_inf_dist}
\end{figure}

Figure \ref{fig:ind_rli_ess} to Figure \ref{fig:ind_overall_inf_dist} show distributions of supply and demand shock-related variables on the industry level. Table \ref{t:industry_details} summarizes a few key statistics for these industries, when further aggregated to 72 industry categories.

{\scriptsize
\begin{longtable}{|p{.8cm}|p{7.75cm}|p{.7cm}|p{.7cm}|p{.9cm}|p{.8cm}|p{.7cm}|p{.8cm}|p{.75cm}|}
  \hline
NAICS & Title & Outp. & Empl. & Demand & Supply & RLI & Essent. & Expos. \\ 
  \hline
111 & Crop Production & 209 & NA & -10 & NA & NA & NA & NA \\ 
  112 & Animal Production and Aquaculture & 191 & NA & -10 & NA & NA & NA & NA \\ 
  113 & Forestry and Logging & 19 & NA & -10 & NA & NA & NA & NA \\ 
  114 & Fishing, Hunting and Trapping & 10 & NA & -10 & NA & NA & NA & NA \\ 
  115 & Support Activities for Agriculture and Forestry & 27 & 378 & -10 & 0 & 14 & 100 & 5 \\ 
  211 & Oil and Gas Extraction & 332 & 141 & -10 & 0 & 47 & 100 & 7 \\ 
  212 & Mining (except Oil and Gas) & 97 & 190 & -10 & -54 & 26 & 27 & 8 \\ 
  213 & Support Activities for Mining & 84 & 321 & -10 & -72 & 28 & 0 & 8 \\ 
  221 & Utilities & 498 & 554 & 0 & 0 & 42 & 100 & 9 \\ 
  23 & Construction & 1636 & 7166 & -10 & -24 & 31 & 66 & 8 \\ 
  311 & Food Manufacturing & 803 & 1598 & -10 & 0 & 21 & 100 & 10 \\ 
  312 & Beverage and Tobacco Product Manufacturing & 192 & 271 & -10 & -3 & 33 & 96 & 7 \\ 
  313-4 & Wholesale Trade & 54 & 226 & -10 & -51 & 26 & 31 & 5 \\ 
  315-6 & Management of Companies and Enterprises & 29 & 140 & -10 & -68 & 25 & 9 & 4 \\ 
  321 & Wood Product Manufacturing & 118 & 402 & -10 & -62 & 26 & 16 & 6 \\ 
  322 & Paper Manufacturing & 189 & 362 & -10 & -8 & 24 & 89 & 6 \\ 
  323 & Printing and Related Support Activities & 80 & 435 & -10 & 0 & 38 & 100 & 4 \\ 
  324 & Petroleum and Coal Products Manufacturing & 618 & 112 & -10 & -26 & 36 & 60 & 7 \\ 
  325 & Chemical Manufacturing & 856 & 828 & -10 & -2 & 38 & 96 & 9 \\ 
  326 & Plastics and Rubber Products Manufacturing & 237 & 722 & -10 & -8 & 28 & 89 & 7 \\ 
  327 & Nonmetallic Mineral Product Manufacturing & 140 & NA & -10 & NA & NA & NA & NA \\ 
  331 & Primary Metal Manufacturing & 239 & 374 & -10 & -73 & 27 & 0 & 7 \\ 
  332 & Fabricated Metal Product Manufacturing & 378 & 1446 & -10 & -59 & 33 & 12 & 6 \\ 
  333 & Machinery Manufacturing & 386 & 1094 & -10 & -49 & 42 & 16 & 5 \\ 
  334 & Computer and Electronic Product Manufacturing & 369 & 1042 & -10 & -38 & 58 & 9 & 4 \\ 
  335 & Electrical Equipment, Appliance, and Component Manufacturing & 132 & 392 & -10 & -31 & 45 & 45 & 6 \\ 
  336 & Transportation Equipment Manufacturing & 1087 & 1671 & -10 & -58 & 37 & 9 & 5 \\ 
  337 & Furniture and Related Product Manufacturing & 77 & 394 & -10 & -47 & 35 & 28 & 4 \\ 
  339 & Miscellaneous Manufacturing & 173 & 601 & -10 & -16 & 40 & 74 & 11 \\ 
  42 & Construction & 1980 & 5798 & -10 & -27 & 50 & 46 & 7 \\ 
  441 & Motor Vehicle and Parts Dealers & 334 & 2006 & -10 & -23 & 43 & 60 & 11 \\ 
  442-4, 446-8, 451, 453-4 & Wholesale Trade & 1052 & 7731 & -10 & -39 & 53 & 17 & 19 \\ 
  445 & Food and Beverage Stores & 244 & 3083 & -10 & -33 & 43 & 43 & 15 \\ 
  452 & General Merchandise Stores & 240 & 3183 & -10 & -49 & 51 & 0 & 17 \\ 
  481 & Air Transportation & 210 & 499 & -67 & 0 & 29 & 100 & 29 \\ 
  482 & Rail Transportation & 77 & 233 & -67 & 0 & 33 & 100 & 11 \\ 
  483 & Water Transportation & 48 & 64 & -67 & 0 & 35 & 100 & 10 \\ 
  484 & Truck Transportation & 346 & 1477 & -67 & -68 & 32 & 0 & 8 \\ 
  485 & Transit and Ground Passenger Transportation & 74 & 495 & -67 & 0 & 27 & 100 & 43 \\ 
  486 & Pipeline Transportation & 49 & 49 & -67 & 0 & 37 & 100 & 8 \\ 
  487-8 & Management of Companies and Enterprises & 146 & 732 & -67 & -10 & 37 & 85 & 8 \\ 
  491 & Postal Service & 58 & 634 & -67 & 0 & 35 & 100 & 10 \\ 
  492 & Couriers and Messengers & 94 & 704 & -67 & 0 & 37 & 100 & 15 \\ 
  493 & Warehousing and Storage & 141 & 1146 & -67 & 0 & 25 & 100 & 6 \\ 
  511 & Publishing Industries (except Internet) & 388 & 726 & 0 & -16 & 70 & 46 & 4 \\ 
  512 & Motion Picture and Sound Recording Industries & 155 & 428 & 0 & -51 & 49 & 0 & 9 \\ 
  515 & Broadcasting (except Internet) & 196 & 270 & 0 & 0 & 65 & 100 & 6 \\ 
  517 & Telecommunications & 695 & NA & 0 & NA & NA & NA & NA \\ 
  518 & Data Processing, Hosting, and Related Services & 207 & 319 & 0 & 0 & 70 & 100 & 4 \\ 
  519 & Other Information Services & 192 & 296 & 0 & -7 & 71 & 75 & 5 \\ 
  521-2 & Construction & 939 & 2643 & 0 & 0 & 74 & 100 & 11 \\ 
  523, 525 & Wholesale Trade & 782 & 945 & 0 & -18 & 74 & 32 & 5 \\ 
  524 & Insurance Carriers and Related Activities & 1231 & 2330 & 0 & 0 & 71 & 100 & 8 \\ 
  531 & Real Estate & 1842 & 1619 & 0 & -53 & 47 & 0 & 19 \\ 
  532 & Rental and Leasing Services & 163 & 556 & 0 & -54 & 46 & 0 & 12 \\ 
  533 & Lessors of Nonfinancial Intangible Assets (except Copyrighted Works) & 182 & 22 & 0 & -30 & 70 & 0 & 7 \\ 
  541 & Professional, Scientific, and Technical Services & 2372 & 9118 & 0 & -2 & 64 & 94 & 9 \\ 
  55 & Management of Companies and Enterprises & 561 & 2373 & 0 & 0 & 66 & 100 & 8 \\ 
  561 & Administrative and Support Services & 971 & 8838 & 0 & -37 & 35 & 44 & 17 \\ 
  562 & Waste Management and Remediation Services & 109 & 427 & 0 & 0 & 30 & 100 & 22 \\ 
  611 & Educational Services & 366 & 13146 & 0 & 0 & 54 & 100 & 29 \\ 
  621 & Ambulatory Health Care Services & 1120 & 7399 & 15 & 0 & 37 & 100 & 59 \\ 
  622 & Hospitals & 933 & 6050 & 15 & 0 & 36 & 100 & 65 \\ 
  623 & Nursing and Residential Care Facilities & 262 & 3343 & 15 & 0 & 28 & 100 & 61 \\ 
  624 & Social Assistance & 222 & 3829 & 15 & 0 & 40 & 100 & 47 \\ 
  711 & Performing Arts, Spectator Sports, and Related Industries & 181 & 505 & -80 & -51 & 44 & 8 & 13 \\ 
  712 & Museums, Historical Sites, and Similar Institutions & 20 & 167 & -80 & -49 & 51 & 0 & 16 \\ 
  713 & Amusement, Gambling, and Recreation Industries & 158 & 1751 & -80 & -65 & 35 & 0 & 21 \\ 
  721 & Accommodation & 282 & 2070 & -80 & -34 & 33 & 50 & 26 \\ 
  722 & Food Services and Drinking Places & 832 & 11802 & -80 & -64 & 36 & 0 & 13 \\ 
  811 & Repair and Maintenance & 235 & 1317 & -5 & -3 & 29 & 96 & 9 \\ 
  812 & Personal and Laundry Services & 211 & 1490 & -5 & -52 & 28 & 28 & 31 \\ 
  813 & Religious, Grantmaking, Civic, Professional, and Similar Organizations & 260 & 1372 & -5 & 0 & 52 & 100 & 19 \\ 
  814 & Private Households & 20 & NA & -5 & NA & NA & NA & NA \\ 
  92 & All Public Sector (custom) & 3889 & 9663 & 0 & 0 & 44 & 100 & 21 \\ 
   \hline
\caption{Key statistics for different 2- and 3-digit NAICS industries. 
Column `Outp.' refers to total output of the industry in current billion USD (2018). 
`Emp.' is total employment in thousands.
`Demand' is the immediate severe demand shock  in \% obtained from the CBO.
`Supply' is the immediate supply shock  in \% derived from the Remote Labor Index and the list of essential industries.
`RLI' is the industry-specific Remote Labor Index  in \%.
`Essent.' is the share of sub-industries being classified as essential  in \%.
`Expos.' denotes the industry-aggregated infection exposure index from O*NET which ranges from 0 to 100 and is explained in Appendix \ref{sec:occupations_at_risk}. 
We make more disaggregated data with further details available in the corresponding data publication.
} 
\label{t:industry_details}
\end{longtable}
}

\section{Occupations most at risk of contracting SARS-CoV-2} \label{sec:occupations_at_risk}
O*NET make available online work context data for occupations that describe the physical and social factors that influence the nature of work. The ``Exposed to disease and infection'' work context\footnote{\url{https://www.onetonline.org/find/descriptor/result/4.C.2.c.1.b?s=2}}, which we refer to as `exposure to infection’ for short, describes the frequency with which a worker in a given occupation is exposed to disease or infection. It ranges from 0 to 100, where 0 means ``never'' and 100 ``everyday''; an exposed to infection rating of 50 means an exposure of "once a month or more but not every week" and 75 means ``Once a week or more but not every day''. We have exposure to infection data for 644 of the 740 occupations in our sample. For those occupations for which we did not have the exposure to infection, when colored them as if they had zero exposure to infection. 

\begin{figure}[H]
    \centering
\includegraphics[width = 0.45\textwidth]{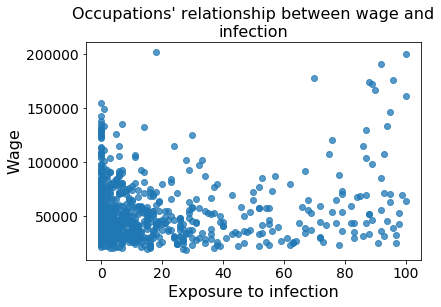}
    \caption{Relationship between wage and probability of infection in linear and log scale. The Pearson correlation is 0.36 (p-value $=1.7\times 10^{-21}$). However we this correlation is mostly driven by high salaries in the health sector, but there are many low-wage occupations with a significant exposure to infection.}
        \label{fig:occ_wage_inf}
\end{figure}

As we see in Figure \ref{fig:occ_wage_inf} there is a U shaped relationship between wages and exposure to infection. There is a correlation of $0.36$ (p-value $=1.7\times 10^{-21}$) between wages and exposure to infection, but this is misleading\footnote{For example, \citet{adams2020}, using survey evidence for $\sim$4000 US individuals, found that workers without paid sick leave are more likely to go to work in close proximity to others, which may have suggested a negative correlation between wages and exposures. Note however that our correlation is based on occupations, not individuals, and that wages are not necessarily an excellent predictor of having paid sick leave or not.}. Though many high wage occupations are highly exposed to infection (high paid doctors), there are also many low wage occupations with high probability of infection. 

\section{Discussion of labor supply shocks which we do not include}

\subsection{Labor supply shocks from mortality and morbidity}
\label{appendix:epidemiology}

\paragraph{Typical estimates.} \citet{mckibbin2020global} consider attack rates (share of population who become sick) in the range 1-30\% and case-fatality rates (share of those infected who die) in the range 2-3\%. From attack rates and case fatality rates, they compute mortality rates. They also assume that sick people stay out of work for 14 days. A third effect they assume is that workers would be care givers to family members.

For their severe scenario of an influenza pandemic, \citet{CBO2006} assumed that 30 percent of the workers in each sector  (except for Farms, which is 10\%) would become ill and would lose 3 weeks or work, at best, or die (2.5\% case fatality rate).

\paragraph{Best guess for current effect of COVID-19.} In the case of COVID-19, estimating a labor supply shock is made difficult by several uncertainties. First of all, at the time of writing there are very large uncertainties on the ascertainment rate (the share of infected people who are registered as confirmed cases), making it difficult to know the actual death rate.

We report the result from a recent and careful study by \citet{verity2020estimates}, who estimated an infection fatality ratio of 0.145\% (0.08-0.32) for people younger than 60, and 3.28\% (1.62-6.18) for people aged 60 or more. The age bracket 60-69, which in many countries will still be part of the labor force, was reported as 1.93\% (1.11–3.89). 

Taking the infection fatality ratio for granted, the next question is the attack rate. In \citet{verity2020estimates}, the infection fatality ratios are roughly one fourth of the case fatality ratios, suggesting that 3/4 of the cases are undetected. For the sake of the argument, consider Italy, a country that has been strongly affected and appears to have reached a peak (at least of a first wave). There are at the time of writing 132,547 cases in Italy\footnote{\url{https://coronavirus.jhu.edu/map.html}} In 2018 the population of Italy\footnote{\url{https://data.worldbank.org/indicator/SP.POP.TOTL?locations=IT}} was 60,431,283. If we assume that Italy is at the peak today and the curve is symmetric, the total number of cases will be double the current number, that is 265,094, which is 0.44\% of the population. If we assume that the true number of cases if 4 times higher, the attack rate is, roughly speaking, 1.76\%. These numbers are more than an order of magnitude smaller than the number who cannot work due to social distancing.

Thus, while it is clear that the virus is causing deep pain and suffering throughout Italy, the actual decrease in labor supply, which is massive, is unlikely to be mostly caused by people being sick, and is much more a result of social distancing measures. 

\paragraph{Uncontrolled epidemic.} Now, it may be informative to consider the case of an uncontrolled epidemic. If we assume that the uncontrolled epidemic has an attack rate of 80\% (a number quoted in \citet{verity2020estimates}), an infection fatality ratio for people in the labor force of 1\% (an arbitrary number between  0.145\% for people younger than 60, and 1.93\% for the 60-69 age bracket) implies an 0.8\% permanent decrease of the labor force. If we assume that those who do not die are out of work for 3 weeks, on an annual basis of 48 worked weeks, we have (3/48)*(0.80-0.01)=4.94\% decrease of the labor supply.

Overall, this exercise suggests that left uncontrolled, the epidemic can have a serious effect on labor supply. However, in the current context, the effect on the economy is vastly more a result of social distancing than direct sickness and death.

\subsection{Labor supply shocks from school closure}
\label{appendix:school}

School closures are a major disruption to the functioning of the economy as parents can no longer count on the school system to care for their children during the day. 

\citet{chen2011social} surveyed households following a school closure in Taiwan during the H1N1 outbreak, and found that 27\% reported workplace absenteism. \citet{lempel2009economic} attempted to estimate the cost of school closure in the US in case of an influenza pandemic. They note that 23\% of all civilian workers live in households with a child under 16 and no stay-at-home adults. Their baseline scenario assumes that around half of these workers will miss some work leading to a loss 10\% of all labor hours in the civilian U.S. economy, for as long as the school closure lasts. 

Some of these effects would already be accounted for in our shocks. For instance, some workers are made redundant because of a supply or demand shock, so while they have to stay at home to care for their children, this is as much a result of labor and supply shock as a result of school closure. For those working from home, we might expect a decline in productivity. Finally, for those in essential industries, it is likely that schools are not close. For instance, in the UK, schools are opened for children of essential workers. Our list of essential industries from Italy includes Education.

Overall, school closure indeed have large effects, but in the current context these may already be accounted for by supply and demand shocks, or non-existent because schools are not fully closed. The loss of productivity from parents working from home remains an open question.

\section{Additional estimates of demand or consumption shocks}
\label{appendix:demand}

In this appendix we provide additional data on the demand shock. Table \ref{table:cbonaics} shows our crosswalk between the Industry classification of the \citet{CBO2006} and NAICS 2-digit industry codes, and, in addition to the ``severe'' shocks used here, shows the CBO's ``mild'' shocks. We have created this concordance table ourselves, by reading the titles of the categories and making a judgement. Whenever NAICS was more detailed, we reported the CBO's numbers in each more fine-grained NAICS.

\begin{table}[ht]
\centering
\begin{tabular}{|c|p{6cm}|p{6cm}|c|c|}
  \hline
NAICS & NAICS & CBO & Severe & Mild \\ 
  \hline
11 & Agriculture, Forestry, Fishing and Hunting & Agriculture & -10 & -3 \\ 
  21 & Mining, Quarrying, and Oil and Gas Extraction & Mining & -10 & -3 \\ 
  22 & Utilities & Utilities & 0 & 0 \\ 
  23 & Construction & Construction & -10 & -3 \\ 
  31 & Manufacturing & Manufacturing & -10 & -3 \\ 
  32 & Manufacturing & Manufacturing & -10 & -3 \\ 
  33 & Manufacturing & Manufacturing & -10 & -3 \\ 
  42 & Wholesale Trade & Wholesale trade & -10 & -3 \\ 
  44 & Retail Trade & Retail trade & -10 & -3 \\ 
  45 & Retail Trade & Retail trade & -10 & -3 \\ 
  48 & Transportation and Warehousing & Transportation and warehousing (including air, rail and transit) & -67 & -17 \\ 
  49 & Transportation and Warehousing & Transportation and warehousing (including air, rail and transit) & -67 & -17 \\ 
  51 & Information & Information (Published, broadcast) & 0 & 0 \\ 
  52 & Finance and Insurance & Finance & 0 & 0 \\ 
  53 & Real Estate and Rental and Leasing & NA & 0 & 0 \\ 
  54 & Professional, Scientific, and Technical Services & Professional and business services & 0 & 0 \\ 
  55 & Management of Companies and Enterprises & NA & 0 & 0 \\ 
  56 & Administrative and Support and Waste Management and Remediation Services & NA & 0 & 0 \\ 
  61 & Educational Services & Education & 0 & 0 \\ 
  62 & Health Care and Social Assistance & Healthcare & 15 & 4 \\ 
  71 & Arts, Entertainment, and Recreation & Arts and recreation & -80 & -20 \\ 
  72 & Accommodation and Food Services & Accommodation/food service & -80 & -20 \\ 
  81 & Other Services (except Public Administration) & Other services except government & -5 & -1 \\ 
  92 & Public Administration (not covered in economic census) & Government & 0 & 0 \\ 
   \hline
\end{tabular}
\caption{Mapping of CBO shocks to NAICS 2-digits} 
\label{table:cbonaics}
\end{table}
\begin{table}[ht]
\centering
\begin{tabular}{|l|cc|}
  \hline
Industry & Consumption shock & Only postponed? \\ 
  \hline
Food, drink, alcohol and tobacco & 0 & NA \\ 
  Clothing and footwear & -50 & yes \\ 
  Housing, heating, etc. & 0 & NA \\ 
  Goods and services (furniture, etc.) & -80 & yes \\ 
  Transport - buying cars & -100 & yes \\ 
  Transport services and car use & -50 & no \\ 
  Recreation and culture - durables & -100 & yes \\ 
  Recreation and culture - games and pets & 0 & NA \\ 
  Recreation and culture - sport and culture & -100 & no \\ 
  Recreation and culture - newspapers and books & 0 & NA \\ 
  Restaurants, hotels and net tourism & -100 & no \\ 
  Miscellaneous (incl health, communication education) & 0 & NA \\ 
   \hline
\end{tabular}
\caption{Demand shock from Keogh-Brown et al. (2010). The first column gives the percentage decrease, while the second column gives the percentage of the first column which will be recouped in future quarters.} 
\label{table:keogh}
\end{table}

\begin{table}[ht]
\centering
\begin{tabular}{|l|c|}
  \hline
Category & Shock (\%) \\ 
  \hline
\multicolumn{2}{|c|}{ISIC.Rev4 shock from OECD (2020)}\\
  \hline
  Manufacturing of transport equipment (29-30) & -100 \\ 
  Construction (VF) & -50 \\ 
  Wholesale and retail trade (VG) & -75 \\ 
  Air transport (V51) & -75 \\ 
  Accommodation and food services (VI) & -75 \\ 
  Real estate services excluding imputed rent (VL-V68A) & -75 \\ 
  Professional service activities (VM) & -50 \\ 
  Arts, entertainment and recreation (VR) & -75 \\ 
  Other service activities (VS)  & -100 \\ 
   \hline
\multicolumn{2}{|c|}{COICOP shock from OECD (2020)}\\
  \hline
  Clothing and footwear (3) & -100 \\ 
  Furnishings and household equipment (5) & -100 \\ 
  Vehicle purchases (7.1) & -100 \\ 
  Operation of private vehicles (7.2) & -50 \\ 
  Transport services (7.3) & -50 \\ 
  Recreation and culture  excluding  package  holidays  (9.1-9.5) & -75 \\ 
  Package  holidays  (9.6) & -100 \\ 
  Hotels  and  restaurants  (11) & -75 \\ 
  Personal  care  services  (12.1)   & -100 \\ 
   \hline
\multicolumn{2}{|c|}{Consumption shocks from Muellbauer (2020)}\\
     \hline
Restaurants and Hotels & -71 \\ 
  Transport services & -70 \\ 
  Recreation services & -63 \\ 
  Food at home & 43 \\ 
  Healthcare & 18 \\ 
   \hline
\end{tabular}
\caption{Estimates of consumption shocks from various sources.} 
\label{table:consumption_shocks}
\end{table}

We also provide two sources of consumption shocks (in principle, these estimates are meant to reflect actual decreases in consumption rather than shifts of the demand curve). Table \ref{table:keogh} shows the consumption shocks used by \citet{keogh2010possible} to model the impact of potential severe influenza outbreak in the UK. Table \ref{table:consumption_shocks} shows the consumption shocks used by \citet{muellbauer2020} to model the impact of the COVID-19 on quarterly US consumption. \citet{OECD2020evaluating} provided two other sources, both reported in Table \ref{table:consumption_shocks}. The first one is based on assumptions of shocks at the industry level, while Table  shows assumptions of shocks by expenditure categories (COICOP: Classification of individual consumption by purpose).


\end{document}